\documentclass[ twocolumn]{aastex62}

\usepackage{natbib}
\usepackage{amssymb, amsmath}
\usepackage{xcolor}
\usepackage{multirow}
 
\usepackage{ifthen}
\usepackage{natbib}
\usepackage{appendix}
\usepackage{etoolbox}
\usepackage[T1]{fontenc}
\usepackage{paralist}
\usepackage{booktabs}
\usepackage{multirow}
\usepackage{float}
\usepackage{url}
\usepackage{hyperref}
\usepackage{soul}
\usepackage{lineno}
\hypersetup{linkcolor=blue,citecolor=blue,filecolor=cyan,urlcolor=blue}

\def\>{\textit{$\textgreater$}}
\def\<{\textit{$\textless$}}

\newcommand{\OIII}{[\rm{O\,\textsc{iii}}]}
\newcommand{\NII}{[\rm{N\,\textsc{ii}}]}
\newcommand{\SII}{[\rm{S\,\textsc{ii}}]}

\def\afrac{\textit{$\rm f_{AGN}$}}
\def\solarm{\text{\rm M$_{\odot}$}}
\def\solarl{\text{\rm L$_{\odot}$}}
\def\lumir{\text{$\rm L_{IR}$}}
\def\lumo{\text{$\rm L_{\OIII}$}}

\def\mass{\text{$\rm M_{*}$}}
\def\kms{\text{$\rm km\ s^{-1}$}}

\def\irac{\emph{The Infrared Astronomical Satellite}}
\def\akari{\emph{AKARI}}
\def\wise{\emph{WISE}}

\shorttitle{{Stellar Kinematics and Ionized gas Outflows in} (U)LIRGs}
\shortauthors{Ayubinia, Xue, Le et al.}

\begin{document}

\title{{Investigation of Stellar Kinematics and Ionized gas Outflows {in Local} (U)LIRGs}}

\author{Ashraf Ayubinia}\thanks{E-mail: ayubinia@mail.ustc.edu.cn}
\affiliation{CAS Key Laboratory for Research in Galaxies and Cosmology, Department of Astronomy, University of Science and Technology of China, Hefei 230026, China}
\affiliation{School of Astronomy and Space Science, University of Science and Technology of China, Hefei 230026, China}
\affiliation{Astronomy Program, Department of Physics and Astronomy, Seoul National University, Seoul 08826, Republic of Korea}

\author{Yongquan Xue}\thanks{E-mail: xuey@ustc.edu.cn}
\affiliation{CAS Key Laboratory for Research in Galaxies and Cosmology, Department of Astronomy, University of Science and Technology of China, Hefei 230026, China}
\affiliation{School of Astronomy and Space Science, University of Science and Technology of China, Hefei 230026, China}

\author{Huynh Anh Nguyen Le}
\affiliation{CAS Key Laboratory for Research in Galaxies and Cosmology, Department of Astronomy, University of Science and Technology of China, Hefei 230026, China}
\affiliation{School of Astronomy and Space Science, University of Science and Technology of China, Hefei 230026, China}

\author{Fan Zou}
\affiliation{Department of Astronomy and Astrophysics, 525 Davey Lab, The Pennsylvania State University, University Park, PA 16802, USA}
\affiliation{Institute for Gravitation and the Cosmos, The Pennsylvania State University, University Park, PA 16802, USA}

\author{Shu Wang}
\affiliation{Astronomy Program, Department of Physics and Astronomy, Seoul National University, Seoul 08826, Republic of Korea}

\author{Zhicheng He}
\affiliation{School of Astronomy and Space Science, University of Science and Technology of China, Hefei 230026, China}

\author{Ece Kilerci}
\affiliation{Sabanc{\i} University, Faculty of Engineering and Natural Sciences, 34956, Istanbul, Turkey}

\begin{abstract}
\noindent
We explore the properties of stellar kinematics and ionized gas in a sample of 1106 local (U)LIRGs from the \akari\ telescope. We combine data from $Wide-field\ Infrared\ Survey\ Explorer$ (\wise)~and Sloan Digital Sky Survey (SDSS) Data Release 13 (DR13) to fit the spectral energy distribution (SED) of each source to constrain the contribution of active galactic nuclei (AGNs) to the total IR luminosity and estimate physical parameters such as stellar mass and star-formation rate (SFR). We split our sample into AGNs and weak/non-AGNs. We find that our sample is considerably above the main sequence. The highest SFRs and stellar masses are associated with ULIRGs. We also fit the H$\beta$ and H$\alpha$ regions to characterize the outflows. We find that the incidence of ionized gas outflows in AGN (U)LIRGs ($\sim$ 72\%) is much higher than that in weak/non-AGN ones ($\sim$ 39\%). The AGN ULIRGs have extreme outflow velocities (up to $\sim$ 2300 km s$^{-1}$) and high mass-outflow rates (up to $\sim$ 60 \solarm~yr$^{-1}$). Our results suggest that starbursts are insufficient to produce such powerful outflows. We explore the correlations of SFR and specific SFR (sSFR) with ionized gas outflows. We find that AGN hosts with the highest SFRs exhibit a negative correlation between outflow velocity and sSFR.  Therefore, in AGNs containing large amounts of gas, the negative feedback scenario might be suggested.
\end{abstract}


\section{introduction} \label{sec:intro}
\irac~(IRAS) was the primary space telescope to carry out an unbiased, sensitive all-sky survey at infrared (IR) wavelengths in the early 1980s (\citealp{Neugebauer1984}), resulting in the detection of particularly interesting extragalactic populations --- (ultra) luminous infrared galaxies ((U)LIRGs) with IR luminosities $\rm L_{\rm 8-1000 \mu m}\geq 10^{11}$ \solarl~for LIRGs and $\rm L_{\rm 8-1000 \mu m}\geq 10^{12}$ \solarl~for ULIRGs (see \citealp{Moorwood1996}; \citealp{Sanders1996}; \citealp{Lonsdale2006} for reviews). {The optical and near-infrared (NIR) imaging analysis of the IRAS 1 Jy sample has demonstrated that nearly all ULIRGs are late-stage mergers (\citealp{Veilleux2002}),} whereas a smaller fraction ($\sim$ 12--25\%) of local LIRGs have interacting morphologies (\citealp{Soifer1984}, \citealp{Hung2014}). The emission from dust grains heated by starburst and/or active galactic nucleus (AGN) activities is the principal source of IR emission from (U)LIRGs because these galaxies contain a great deal of dust that absorbs the UV/optical emission of young stars and AGNs and  then reradiates thermal IR emission ({see, e.g., \citealp{Lonsdale2006} for a review}). The coexistence of starbursts and AGNs in (U)LIRGs makes these populations indispensable to understanding the coevolution of the central supermassive black holes (SMBHs) and their host galaxies. It is widely believed that they are a key phase of galaxy evolution. {For instance, based on the similarity between the optical/NIR spectral energy distributions (SEDs) of massive extremely red galaxies (ERGs) that are ULIRGs and those that are not (albeit with similar stellar masses and redshifts), \citet{Caputi2006} argued that ULIRGs are not associated with a particular type of NIR-selected galaxies, but rather a phase (the phase of high star-formation/quasar activity) during the life of a massive galaxy.} Nevertheless, their physical nature is still not well understood due to their limited number in the local universe (e.g., \citealp{Kim1998}) and their short lifespans (e.g., \citealp{Vernet2001}).

{{From the observational point of view, both stellar feedback (\citealp{Heckman2000}; \citealp{Pettini2000}; \citealp{Ohyama2002}) and AGN feedback (\citealp{Lynds1967}; \citealp{Feruglio2010}; \citealp{Fabian2012}; \citealp{King2015}) are crucial in regulating star formation and SMBH growth through the injection of energy into the interstellar medium (ISM) as required by the cosmological simulations (e.g., \citealp{Taylor2017}; \citealp{Li2018}). The galactic-scale outflows (see the review by \citealp{Veilleux2002}) are believed to be the primary form of feedback {(e.g., \citealp{Chen2022}; \citealp{He2022}).}}}\

Despite the {complex nature of (U)LIRGs}, which usually exhibit powerful ionized (e.g., \citealp{Rupke2013a}; \citealp{Rich2011}; \citealp{Rich2014}; \citealp{Toba2017}; \citealp{Baron2020}), neutral atomic (e.g., \citealp{Rupke2005}, \citealp{Rupke2013a}; \citealp{Cazzoli2016};  \citealp{Baron2020}) and molecular (e.g., \citealp{Rupke2013b}; \citealp{Sturm2011}; \citealp{Alfonso2017}; \citealp{Falstad2019}; \citealp{Imanishi2019}) gas outflows along with relatively high star-formation rates (SFRs) (e.g., \citealp{Malek2017}), it is essential to figure out {their} primary power source and understand the relationship between coexisting AGN feedback and stellar feedback.

There have been several attempts to demonstrate how the properties of multiphase outflows in dust-enshrouded starbursts differ from those in AGN (U)LIRGs. \citet{Rupke2005} investigated the cool gas component via the NaID absorption doublet ($\lambda \lambda 5890,5896$ {\AA}) features in a sample of low-$z$ starbursts and Seyfert 2 ULIRGs. They found that distinguishing neutral gas outflow properties between starbursts and AGNs is challenging; however, for fixed SFRs, AGN ULIRGs have higher neutral gas outflow velocities, indicating a significant AGN contribution to driving outflows (see also, e.g., \citealp{Sturm2011}). The optical spectroscopic studies also suggest that the powerful ionized winds in AGN ULIRGs, in which star formation is extremely active, are AGN-driven (e.g., \citealp{Rodriguez2013}). Thanks to the advent of millimeter/submillimeter telescopes such as the Atacama Large Millimeter Array (ALMA), high-spatial-resolution images of molecular gas in nearby galaxies have enabled us to reveal starburst-driven (e.g., \citealp{Salak2020}) and AGN-driven (e.g., \citealp{Spoon2013}) molecular outflows in more detail. However, it remains particularly challenging to determine the origin of outflows in (U)LIRGs because of their heavy obscuration and the lack of high-sensitivity, spatially-resolved observations for a large enough number of (U)LIRGs to disentangle their complex kinematics and outflow properties.

Moreover, the evolutionary track of (U)LIRGs across cosmic time still remains inconclusive. It is expected that because of the rapid growth of central SMBHs and feedback processes, early-stage starburst (U)LIRGs with younger stellar populations transform into late-stage luminous AGN (U)LIRGs (e.g., \citealp{Netzer2007}; \citealp{Hou2011}). To uncover the mysterious veil of (U)LIRGs and to understand their evolutionary stages, highly reliable and fairly complete samples are required.

This study presents a systematic investigation of the stellar kinematics and ionized gas outflow properties of the heretofore largest sample of low-redshift (U)LIRGs assembled. This paper is organized as follows. In Section \ref{sec:sample}, we describe the construction of our sample. The models used to fit the (U)LIRG SEDs, as well as their optical {spectral} fitting to detect ionized gas outflows, are shown in Section \ref{sec:analysis}. In Section \ref{sec:discussion}, we discuss the properties of ionized gas outflows and their correlations with galaxy properties. We summarize our results in Section \ref{sec:results}. 

Throughout the paper, we use {a cosmology of} $\rm H_0 = 70 \ km\ s^{-1}\ Mpc^{-1}$, $\rm \Omega_M = 0.27$, and $\rm \Omega_\Lambda = 0.73$,  and adopt rest-frame wavelengths for the analyses.

\section{sample construction} \label{sec:sample}
Our {(U)LIRG} sample is selected from the IR {galaxy catalog} provided by \citet{Kilerci2018}. Briefly, they cross-matched (see Section 2 {of} \citealp{Kilerci2018} for their criteria) the \akari/FIS all-sky survey bright source {catalog (version 2)}\footnote{\url{https://www.ir.isas.jaxa.jp/AKARI/Archive/Catalogues/FISBSCv2/}}, which provides the positions and fluxes at 65, 90, 140 and 160 $\mu$m for 918,056 sources, with {the} \akari/IRC all-sky survey source {catalog (version 1)}\footnote{\url{https://www.darts.isas.jaxa.jp/astro/akari/data/index.html}} and the {\it Wide-field Infrared Survey Explorer} (\wise)/AllWISE Source catalog\footnote{\url{https://wise2.ipac.caltech.edu/docs/release/allwise/}} to include mid-infrared (MIR) data when available. Subsequently, by combining the data from Sloan Digital Sky Survey (SDSS) Data Release 13 (DR13, \citealp{Albareti2017}), 6-degree Field Galaxy Survey (6dFGS, \citealp{Jones2004};\citealp{Jones2005};\citealp{Jones2009}) and the 2MASS Redshift Survey (2MRS, \citealp{Huchra2012}), they measured total IR {luminosities} integrated over 8--1000 $\mu$m for more than 15,500 IR galaxies at redshifts of $z \leq$ 0.3 using the photometric redshift code, Le PHARE.\footnote{\url{https://www.cfht.hawaii.edu/\~arnouts/LEPHARE/lephare.html}} 

In this work, we first define a parent sample of 4705 local IR galaxies from this catalog with available optical counterparts in SDSS and {extinction-corrected Petrosian $r$ magnitude $ r_{petro} \leq 17.7$ (see Section 2.2.1 of \citealp{Kilerci2018} for details). Based on this optical cut, our current sample does not include very faint (U)LIRGs. However, this will not significantly affect our final results because the optical spectra of these faint sources typically lack sufficient signal-to-noise ratios (S/Ns) or are even unavailable.
Then, we select objects with $\lumir \geq 10^{11}$ \solarl~({i.e., (U)LIRGs) from the parent sample. Almost all of our selected sources are detected in far-inferared (FIR) wavelengths, 100\% and 95\% in 90 $\mu$m and 65 $\mu$m (around the peaks of (U)LIRG SEDs), respectively. To model the SEDs of our sources reliably, we further exclude 94 sources with no MIR data, no detection in the \wise~[3.4]--[12] $\mu$m and/or in  \akari~9--18 $\mu$m. Consequently, our main sample consists of 1106 (U)LIRGs at $z~\<$ 0.3 with reliable spectroscopic redshifts and photometric measurements, including 1066 LIRGs, heretofore being the largest sample of LIRGs that have been less explored than ULIRGs (see \citealp{Stierwalt2013}; \citealp{Armus2009} for examples of already large samples of LIRGs). Figure \ref{fig:redshift} illustrates IR luminosity as a function of redshift for our parent and main samples. 

\begin{figure}
\includegraphics[width=.95\linewidth]{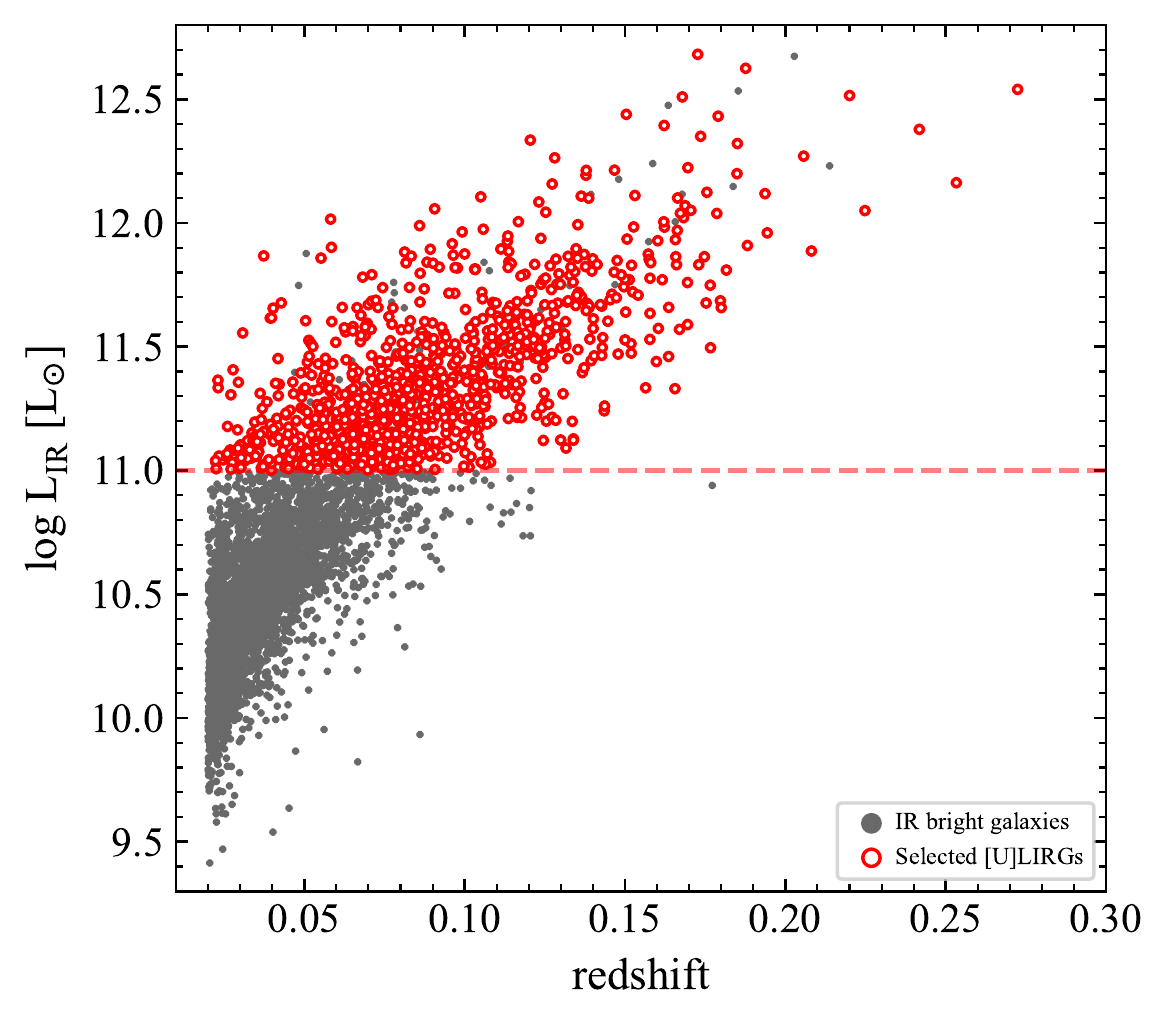}
\caption{Total IR luminosity (integrated over 8--1000 $\mu$m) versus redshift for our parent (gray) and main (red) samples. The red dashed line shows \lumir = $10^{11}$ \solarl.}
\label{fig:redshift}
\end{figure}

\section{{Analyses and Results}}\label{sec:analysis} 
\subsection{{SED} Fitting}\label{sec:sed}
Thanks to the FIR and MIR data, as well as the optical data, we are able to model the optical-IR part of our (U)LIRG SEDs. To do so, we apply the intrinsic extinction correction for SDSS magnitudes (in the bandpasses $u, g, r, i$, and $z$) using the dust map of \citet{Schlafly2011} and the extinction law of \citet{Fitzpatrick1999}. After correcting for the zero-point offsets between the SDSS and AB systems, we convert magnitudes into flux densities. The \wise~[3.4] and [4.6] $\mu$m magnitudes are corrected for extinction using the dust map of \citet{Schlafly2011} and the extinction law of \citet{Flaherty2007}. Next, the four \wise~magnitudes are converted into flux densities using the zero-magnitude flux densities and color corrections of \citet{Wright2010}. By combining the \akari~photometric data, we have a maximum of 14 data points for SED-fitting.

To perform SED-fitting in a self-consistent model framework, we employ the CIGALE SED modelling code\footnote{\url{https://cigale.lam.fr}} (\citealp{Boquien2019}), which relies on the energy balance principle (i.e., the conservation of energy between the UV-optical emission absorbed by dust and MIR-FIR emission re-radiated by the same dust). Briefly, CIGALE constructs a high-dimensional parameter grid of SED models and identifies the best fit by minimizing the $\chi^2$ statistic. Many studies have previously utilized CIGALE to constrain the physical parameters of (U)LIRGs (e.g., \citealp{Malek2017}; \citealp{Vika2017}; \citealp{Dietrich2018}).

We adopt a delayed star-formation history (SFH) as SFR $\rm \varpropto t \times exp(-t/\tau$) with an exponential burst of star formation that has been ongoing during the past 5--1000 Myr. Since our sources are at low redshifts, it is appropriate to consider the old stellar populations as old as 10 Gyr. We rely on the initial mass function (IMF) of \citet{Salpeter1955} and the stellar population synthesis models provided by \citet{Maraston2005} with solar metallicity. This model includes the contribution from the thermally pulsing asymptotic giant branch (TP-AGB) stars that is essential for modeling young stellar populations (\citealp{Maraston2005}). Most (U)LIRGs are dust-obscured, and this dust absorbs the UV-to-NIR radiation from stars and AGNs very effectively and re-emits it in MIR and FIR wavelengths. Therefore, it is vital to model dust attenuation properly. We model attenuation curves with the attenuation laws inspired from \citet{Calzetti2000}, allowing color excess E(B$-$V) of the young population to vary in a wide range of values. Due to less dust enshrouding, the old stellar population suffers less reddening than the young stellar population. Thus, a reduction factor is essentially defined as $\rm E(B-V)_{old}/ E(B-V)_{young}$. We apply three different reduction factors to dictate the quantity of the old stellar population attenuation. \

To fit the reprocessed IR radiation from dust, the templates of \citet{Dale2014} are employed. The distribution of dust emission from \citet{Dale2014} is described by a power-law model as $\rm d M_{d} \propto  U^{-\alpha} dU$, with $\rm M_{d}$ as the dust mass heated by a radiation field at intensity U. 
We allow the IR power-law slope ($\alpha$) to vary between 0.0625 and 2.5, covering a wide range of activities from starbursts to quiescent galaxies.\

 It is expected that a fraction of our (U)LIRGs require an AGN component to fit their observed SEDs. In the updated version of CIGALE (i.e., X-CIGALE, \citealp{Yang2020}), there are two AGN models: \citet{Fritz2006} and SKIRTOR (\citealp{Stalevski2017}). The former is built upon radiative-transfer modeling for a smooth torus, while the latter considers a clumpy torus. Most recently, \citet{Ramos2022} showed no significant difference in the fractional contribution of AGNs to the total IR luminosity (\afrac) estimated from either a smooth or clumpy torus. Therefore, we model AGN emission using models presented in \citet{Fritz2006}, which consider two main components: the central-source emission and the circumnuclear dusty torus emission. A set of seven parameters is used to determine the model: the ratio of the outer to the inner radius of the dusty torus (r), the optical depth at 9.7 $\mu$m ($\rm \tau_{9.7}$), the two parameters ($\rm \beta, \gamma$) that describe the dust density distribution ($\rm \propto r^{\beta}e^{-\gamma|cos\theta|}$), the opening angle of the torus ($\rm \theta$), the angle between the torus equatorial plane and the line of sight ($\rm \psi$), and the fractional contribution of AGNs to the total IR luminosity (\afrac). We choose three input values for $\rm \tau_{9.7}$; the low ($\rm \tau_{9.7}$ = 1), medium ($\rm \tau_{9.7}$ = 3) and high ($\rm \tau_{9.7}$ =  6) optical depth models. It has been shown in detail that the SED-fitting and the estimation of physical parameters such as SFR or \afrac~are insensitive to viewing angles as long as both Type 1 and Type 2 representatives are included (\citealp{Ramos2022}). Hence, $\rm \psi$ is set to be 0.001$^{\circ}$ and 89.99$^{\circ}$ to include both Type 1 and Type 2 AGNs, respectively. We allow \afrac~to vary between 0.0 and 0.8. Other parameters are fixed to the values found by \citet{Fritz2006}. More details of SED-fitting parameters are presented in Table~\ref{tab:cigale}. 

\begin{table*}
        \begin{center} 
                \caption{Modules and parameter ranges used for the SED-fitting with CIGALE.}
                \label{tab:cigale}
                \begin{tabular}{p{0.42\linewidth}p{0.35\linewidth}} \hline \hline
                	   Parameter &  Range \\
		       \hline 
                      \multicolumn{2}{c}{{Delayed} SFH with an Exponential Burst} \\  \hline  
                        e-folding time of the main stellar population (Myr) & 500, 1000, 3000, 5000, 7000\\
                        e-folding time of the late starburst population (Myr) & 300\\
                        mass fraction of the late burst population & 0.001, 0.005, 0.01, 0.05, 0.1, 0.5\\
                        age of the late burst (Myr) & 5, 10, 25, 50, 100, 200, 350, 500, 750, 1000\\
                        \hline
                        \multicolumn{2}{c}{{Stellar Population Synthesis} (\citealp{Maraston2005}) } \\  \hline
                        initial mass function & 0 (\citealp{Salpeter1955})\\
                        {metallicity} (solar metallicity) & 0.02 \\
                        \hline
                        \multicolumn{2}{c}{Attenuation Law (\citealp{Calzetti2000})} \\ \hline
                        {color} excess of the young stellar population & 0.005, 0.01, 0.05, 0.15, 0.3, 0.45, 0.6, 0.75, 0.9, 1.1, 1.4, 1.7, 2.0, 2.3\\
                        reduction factor for the E(B$-$V) of the old population &  0.25, 0.5, 0.75\\ 
                        slope correction of the power law & 0.0\\
                        \hline
                        \multicolumn{2}{c}{Dust Emission (\citealp{Dale2014})} \\   \hline 
                         IR power-law slope & 0.0625, 0.5, 1.0, 1.5, 2.0, 2.5 \\                 
              	       \hline
	       	        \multicolumn{2}{c}{AGN Emission (\citealp{Fritz2006} )} \\   \hline 
                         optical depth at 9.7 micron & 1.0, 3.0, 6.0 \\ 
                         opening angel of the dusty torus (deg) & 100.0\\     
                         angle between equatorial axis and {line of sight} (deg)& 0.001, 89.990\\     
                         fractional contribution of AGN to total \lumir & 0.0, 0.01, 0.02, 0.04, 0.06, 0.08, 0.1, 0.12, 0.14, 0.16, 0.18, 0.2, 0.23, 0.25, 0.28, 0.3, 0.33, 0.35, 0.38, 0.4, 0.45, 0.5, 0.55, 0.6, 0.65, 0.7, 0.75, 0.8\\   
                         \hline  
                     
                \end{tabular} 
        \end{center}
\end{table*}

We choose the \texttt{pdf\char`_analysis}~(PDF: probability distribution function) module in CIGALE to estimate various physical properties of our {sources} from likelihood-weighted parameters on a fixed grid of models. The threshold for an acceptable fit is the best-fit reduced chi-square {($\chi_{\rm red}^{2}) \leq$ 5}. By applying this criterion, 116 sources are excluded from our analyses due to unreliable SED-fitting results, and 990 sources remain for further investigation. The median value of $\chi^{2}$ for accepted fits is 1.63. 
Moreover, the comparison between true values and mock analysis is also widely adopted for assessing the reliability of SED-fitting results (e.g., \citealp{Ciesla2015}; \citealp{Malek2017}; \citealp{Boquien2019}). In brief, the mock observations for each target are built by adding noise to the best-fit SED model and convolving it into the set of filters. The noise is taken randomly from a Gaussian distribution with the same standard deviation as the uncertainty on the observation (\citealp{Boquien2019}). Hence, we examine the reliability of our accepted SED-fitting results by comparing true and mock values. As shown in Figure \ref{fig:mock}, there is good agreement between true and mock values.\

There are a couple of approaches to identifying AGNs in a sample of galaxies. In CIGALE, the AGN fraction is defined as {\afrac~= \lumir(AGN)/\lumir, where \lumir(AGN) is the IR luminosity contributed from the AGN and \lumir~is the total IR luminosity. The \afrac~$\sim$ 0\% indicates that star-formation activities dominate IR emission, {while} the \afrac~$\sim$ 100\% suggests that IR emission is entirely from the AGN. Nonetheless, the value of this parameter cannot necessarily quantify the AGN contribution properly, especially when the SFR is high or the AGN contribution is negligible. For example, \citet{Ciesla2015} showed that below \afrac~= 10\%, the AGN contribution is overestimated significantly, and the fractional difference between the values derived from CIGALE and the simulated ones} can rise up to $\sim$ 120\%, regardless of AGN SED shape (Type 1 or Type 2). They also found that when the AGN contribution is low (\afrac~$\sim$ 5\%), CIGALE is unable to distinguish between an SED with a small AGN contamination and an SED without an AGN contribution. As an alternative, the BPT diagnostic diagram (\citealp{Baldwin1981}) has been widely used to distinguish star-forming galaxies and AGNs based on the optical emission-line ratios, which are, however, only usable when reliable measurements of emission-line fluxes are available. Instead, it has been proven that \wise~colors are an effective tool that is able to select AGNs very successfully (e.g., \citealp{Stern2012}; \citealp{Assef2013}). Therefore, we split our sample into AGNs and weak/non-AGNs based on a simple MIR color criterion, \wise~[3.4]--[4.6] $\mu$m $\geq$ 0.8 (Vega magnitude) for AGNs (\citealp{Stern2012}) which enables us to classify both obscured and unobscured AGN (U)LIRGs. Figure \ref{fig:wise} shows the \wise~color-color diagram for the 908 sources detected in all \wise~[3.4]--[12] $\mu$m bands compared with the AGN fraction parameter from SED-fitting. Accordingly, we identify 119 AGNs and 789 weak/non-AGNs. We mainly select AGNs based on the \wise~MIR criterion; however, among the 82 sources in our sample that are not detected in \wise~(at least in one of the [3.4]--[12] $\mu$m bands), there may still be some AGNs left. Therefore, we try to supplement the AGN sample through the SED method. Consequently, we employ the \afrac~parameter to classify them into AGNs and weak/non-AGNs when \afrac~$\geq$ 10\% and \afrac~< 10\%, respectively, which is overall reasonable, as shown in Figure~\ref{fig:wise}. Based on this method, we select 22 AGNs (including 3 ULIRGs) and 60 weak/non-AGNs (with no ULIRGs). We reiterate that the SED-based AGN sample is much smaller than the original WISE-based AGN sample; only $\sim$ 16\% (22/141) of our AGNs are classified based on AGN fraction, and it does not affect our final results materially.

\begin{figure}
\includegraphics[width=.95\linewidth]{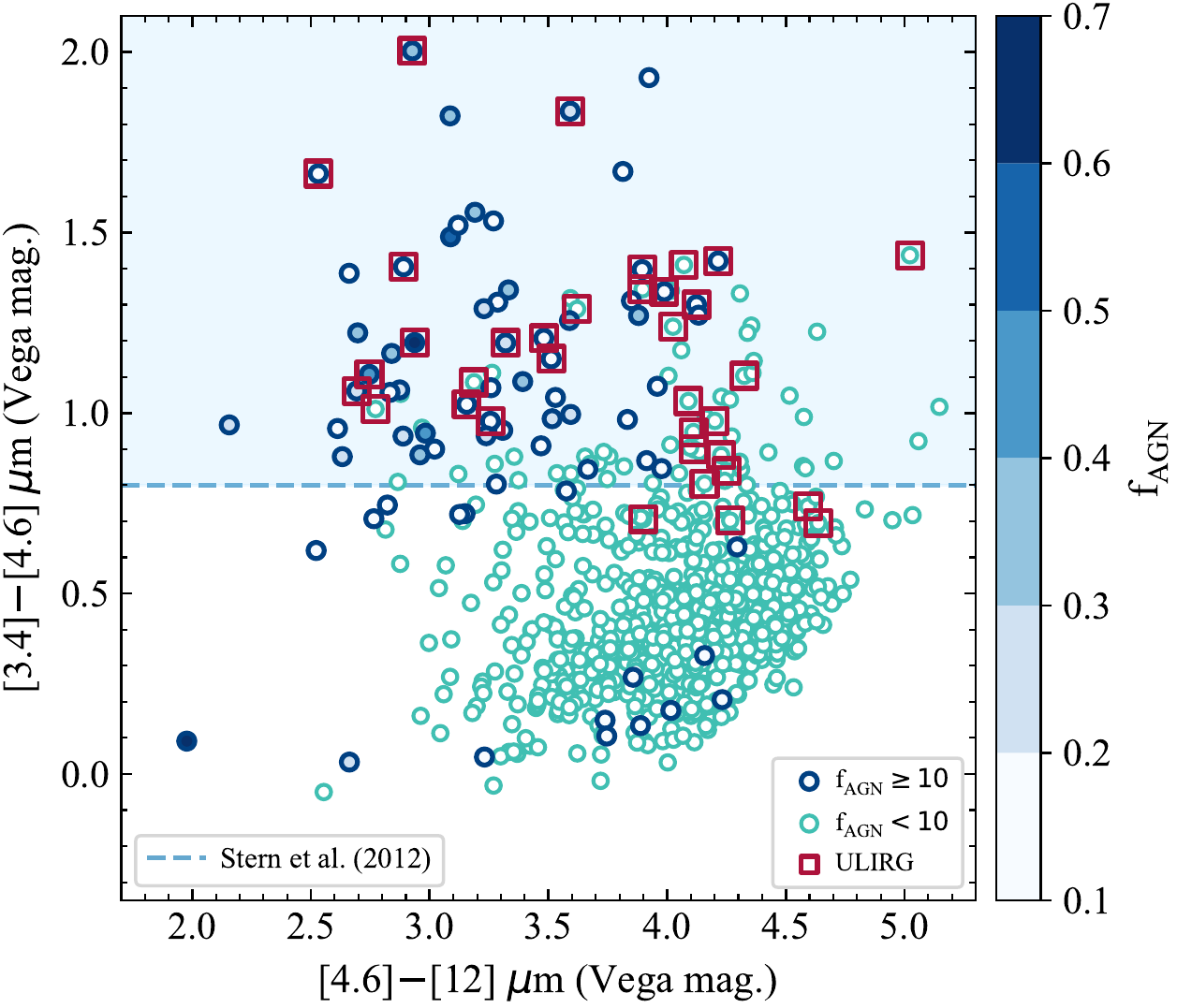}
\caption{\wise~color-color diagram with color-coded \afrac~from SED-fitting for our \wise-detected (U)LIRGs. The blue and green circles {represent} sources with \afrac~$\geq$ 10\% and \afrac~< 10\%, respectively. The red squares indicate ULIRGs. The dashed line and the shaded region above indicate the AGN-selection criterion from \citet{Stern2012}.}
\label{fig:wise}
\end{figure}

The combination of the \wise-color and \afrac-based selections provides us with the identification of 141 AGNs and 849 weak/non-AGNs in our sample of 990 (U)LIRGs with {acceptable SED-fitting results ($\chi_{\rm red}^{2} \leq$ 5)}. Different studies, which mostly focused on ULIRGs, reported different percentages of local (U)LIRGs hosting an AGN (e.g., \citealp{Farrah2007}; \citealp{U2012}; \citealp{Ichikawa2014}), finding a local AGN ULIRG fraction of $\sim$ 25--70\%, depending on IR luminosity (e.g., \citealp{Veilleux1997};  \citealp{Veilleux2002}; \citealp{Nardini2010}). Here, we find that only a small fraction ($\sim$ 14\%) of our (U)LIRGs can be identified as AGNs. We note that the vast majority of our sources are LIRGs (rather than ULIRGs) which usually have smaller AGN contributions (\citealp{Arribas2014}). Incidentally, almost all of our ULIRGs are associated with AGNs (34/38), although some of them have \afrac~< 10\% (see Figure  \ref{fig:wise}). As mentioned above, \afrac~is not always a good indicator of the level of AGN contribution; especially when the SFR is high (see Figure \ref{fig:sfr_mass}), a small {\afrac}~may not be well constrained.\

The relation between SFR and stellar mass (\mass) is presented in Figure \ref{fig:sfr_mass}. For comparison purposes, we also present local samples of (U)LIRGs from previous studies (\citealp{Pereira-Santaella2015}; \citealp{Malek2017}; \citealp{Jarvis2020}) in which SFR and \mass~are driven from SED-fitting process. As expected, the vast majority of our sources are significantly above the main sequences ($\sigma = 0.42$ dex) found by \citet{Elbaz2007} and \citet{Aird2017} at $z \sim$ 0 and $z \sim$ 0.3, respectively, indicative of vigorous star formation in our (U)LIRGs. Moreover, our sample is also above the threshold SFR of LIRGs to reach a \lumir~> 10$^{11}$ \solarl~found by \citet{Jarvis2020} using the correlation between \lumir~due to star formation and SFR. Some previous non-(U)LIRG studies (e.g., \citealp{Xue2010}; \citealp{Guo2020}) found that AGNs tend to reside in massive galaxies. In our sample, although the highest stellar masses are associated with AGN ULIRGs, we find no statistically significant difference in \mass~distribution between AGN and weak/non-AGN (U)LIRGs, according to the Kolmogorov-Smirnov test (KS test; $p$-value = 0.06). \citet{Aird2012} showed that the prevalence of AGNs in massive galaxies might be caused by the sensitivity of sample selection to the underlying Eddington ratio distribution; AGNs can be hosted by galaxies with a wide range of \mass. {Interestingly}, we find that AGN (U)LIRGs {tend to have higher SFRs (median log SFR/(\solarm~yr$^{-1}$) = 1.94) than weak/non-AGN (U)LIRGs (median log SFR/(\solarm~yr$^{-1}$) = 1.62), with a KS test $p$-value = $1.48 \times 10^{-11}$}. Elevated SFRs in AGN hosts have been reported compared to {weak/non-AGN star-forming} and quiescent galaxies in $Herschel$-detected samples (e.g., \citealp{Santini2012}). It may suggest that SFRs of host galaxies are enhanced by AGN activity in our AGN (U)LIRGs. Clearly, ULIRGs {tend} to have both higher SFRs and \mass~in comparison with LIRGs (see Figure \ref{fig:sfr_mass}). This result is consistent with the expected relationship between SFR and dust luminosity, as SFR increases with dust luminosity (\citealp{Malek2017}). ULIRGs, which are often found in interacting or merging galaxies, usually have much more dust and gas than LIRGs.\

\begin{figure}
\includegraphics[width=.95\linewidth]{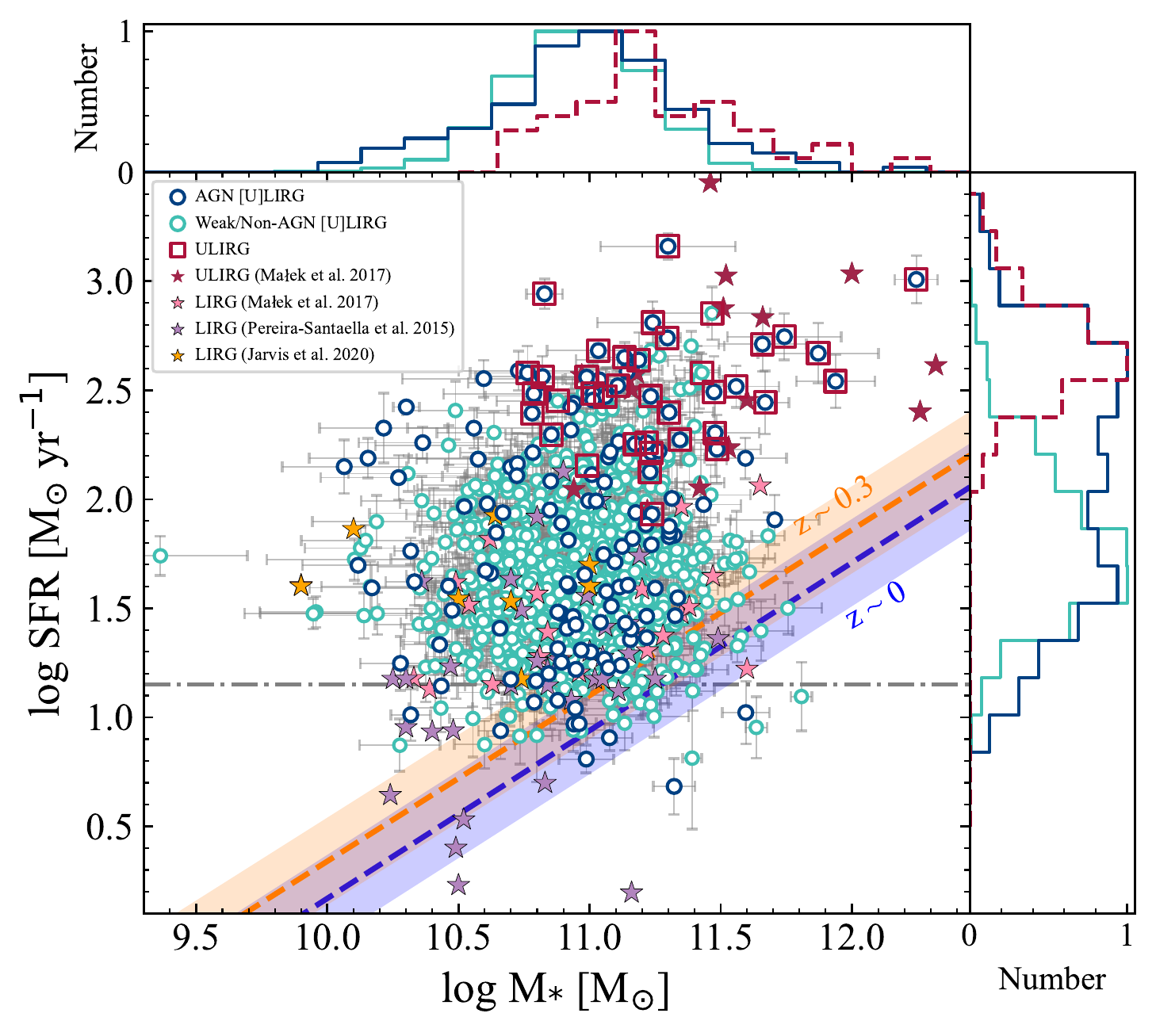}
\caption{Relationship between SFR and stellar-mass for our AGN (blue circles) and weak/non-AGN (green circles) (U)LIRGs. ULIRGs are presented in red squares. The blue and orange dashed lines show the main sequences at $z \sim$ 0 (\citealp{Elbaz2007}) and $z \sim$ 0.3 (\citealp{Aird2017}), respectively, with the shaded areas indicating 0.2-dex uncertainties. The gray dashed–dotted line presents the limit for LIRGs (the threshold SFR to reach an \lumir~>  10$^{11}$ \solarl)  obtained by \citealp{Jarvis2020}. Stars indicate (U)LIRG samples from \citealp{Pereira-Santaella2015}, \citealp{Malek2017} and \citealp{Jarvis2020}. The histograms show the normalized distributions of SFR and stellar-mass.}
\label{fig:sfr_mass}
\end{figure}

\subsection{Optical Spectral Fitting}\label{sec:spectra}
\subsubsection{Stellar Velocity Measurement}\label{sec:vel_dispersion}
{To estimate the central stellar velocity dispersions ($\sigma_{*}$) and systemic velocity from the stellar absorption lines}, we use the Penalized Pixel-Fitting code, version 6.7.16 (pPXF\footnote{\url{https://www-astro.physics.ox.ac.uk/\~mxc/software/\#ppxf}}: \citealp{Cappellari2004}; \citealp{Cappellari2017}). The pPXF code convolves the stellar population templates by a parameterized line-of-sight velocity distribution (LOSVD) to create a model galaxy spectrum $\rm G_{mod}$ (x). Then, the best-fit parameters of the LOSVD are determined by $\chi^2$ minimization of fitting the $\rm G_{mod}$ (x) to an observed galaxy spectrum. For spectral templates, we use 47 empirical stellar templates with solar metallicity from the MILES library (\citealp{Sanchez2006}), an IMF slope of 1.3 and ages of 0.06--12.6 Gyr. We fit the spectra in the rest-frame wavelength range of 3800--5400 \AA. The minimum and maximum S/Ns of our source spectra are 7 and 109, respectively, with a median of 22. We achieve a good estimation of velocity dispersion of our (U)LIRGs, with $\Delta\sigma_{*}$~(error of the stellar velocity dispersion) $\<$ 40 km s$^{-1}$ in 86\% of our sample. The distribution of velocity dispersion of our sample peaks in the range of 100--200 km s$^{-1}$. \

{The pPXF results of 91/990 sources with unreliable measurements, either having $\Delta\sigma_{*}/\sigma_{*}~ \>~1$ or with their stellar continua and absorption lines not being visually adequately fitted, are not included in the following analyses.}
The median values of stellar velocity dispersion for AGN and weak/non-AGN (U)LIRGs are {155 and 145} \kms, respectively. The result of the KS test suggests that the stellar velocity distributions of AGN and weak/non-AGN (U)LIRGs are not significantly different {($p$-value = 0.017)}. In Figure \ref{fig:sigma_mass}, we present the distribution of $\sigma_{*}$ as a function of \mass. Since our sample spans a narrow range of stellar mass, we do not try to fit \mass--$\sigma_{*}$ relation, rather; we compare the position of our targets relative to the best-fit of \mass--$\sigma_{*}$ relation given by \citet{Woo2017} using a large sample of Type 2 AGNs (presented as the dashed line). We  also overlap four samples of (U)LIRGs with available measurements of stellar velocity dispersion and mass from the literature. We note that for ULIRGs adopted from \citet{Colina2005} and \citet{Rothberg2013}, we assume that the dynamical mass of targets is equal to their stellar mass. For all but two targets in \citet{Medling2014}, the stellar mass measurements provided by \citet{U2012}. For two sources, we use dynamical mass (\citealp{Medling2014}. For all six LIRGs in \citet{Crespo2021}, we adopt the stellar mass calculated by \citet{Pereira-Santaella2015}. In general, it seems that (U)LIRGs follow \mass-- $\sigma_{*}$ relation, however, with some deviation. Our sample tends to be slightly located below the \mass-- $\sigma_{*}$ relation given by \citet{Woo2017}. It can be due to this reason that the extraction of the stellar kinematics of (U)LIRGs using 1D SDSS spectra obtained at only a single point is difficult. Unlike ULIRGs, which very often exhibit merging morphology, LIRGs are detected in a wide range of morphologies such as isolated disks, disturbed spirals, or (early/late stage) mergers (e.g., \citealp{Petric2011}), and they, therefore, present complex rotating disk structure (e.g., \citealp{Crespo2021}). Hence, the velocity dispersion measurements may not be well constrained, especially for targets whose spectra have low S/Ns (e.g., \citealp{Sexton2020}).

\begin{figure}  
\includegraphics[width=.95\linewidth]{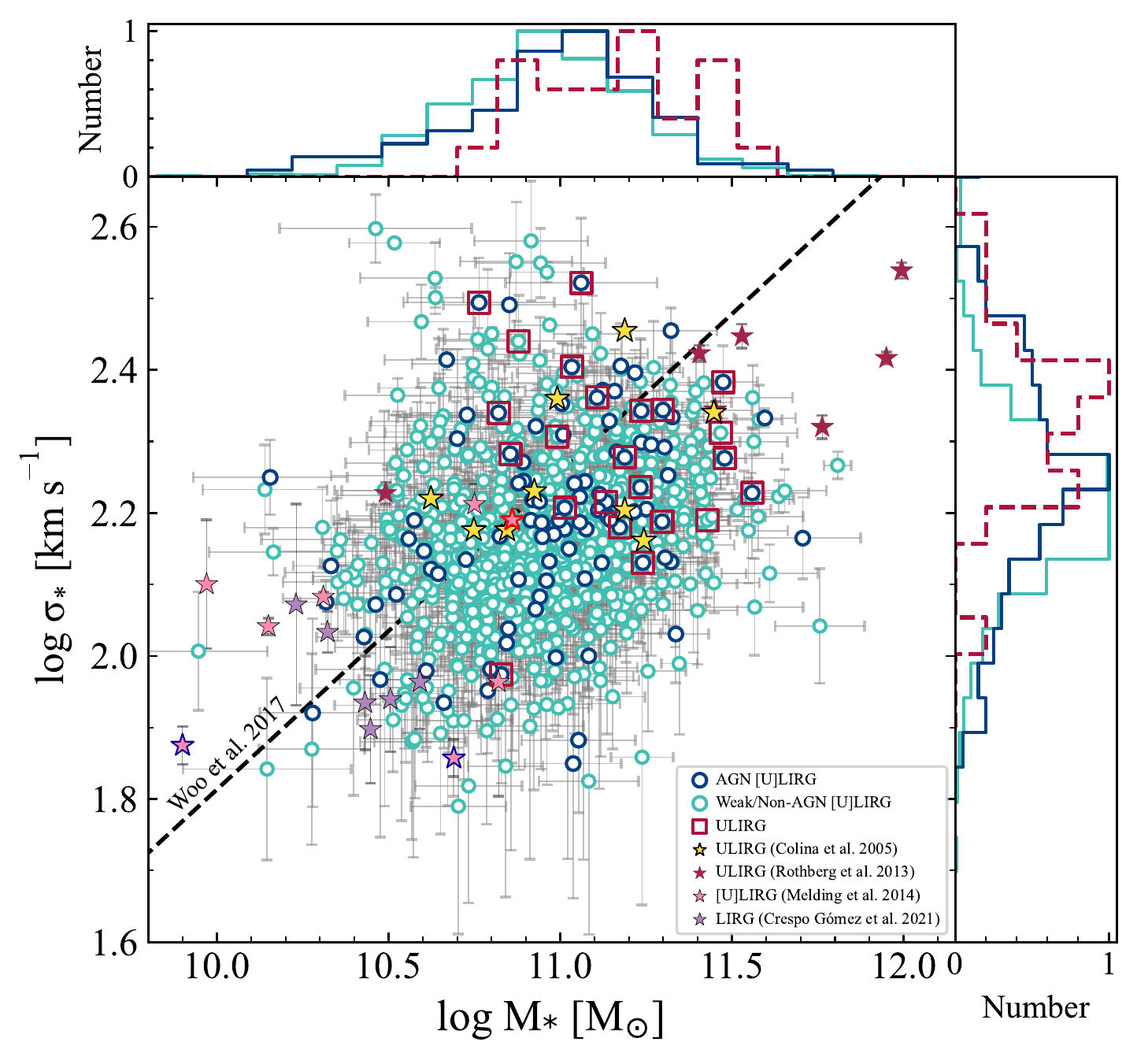}
\caption{Stellar velocity dispersion as a function of stellar mass. Our AGNs, weak/non-AGNs and ULIRGs are presented in the same colors as in Figure \ref{fig:sfr_mass}. Stars indicate (U)LIRG samples from \citet{Colina2005}, \citet{Rothberg2013}, \citet{Medling2014} and \citet{Crespo2021}. The star with red outline color presents an ULIRG, while two stars with blue outline color indicate two sources with dynamical mass from \citet{Medling2014}. The histograms show the normalized distributions of stellar velocity dispersion and stellar-mass.}
\label{fig:sigma_mass}
\end{figure}

\subsubsection{Detection of Ionized Gas Outflows}\label{sec:outflows}
To identify ionized gas outflows in our (U)LIRGs, we fit their optical spectra with the multicomponent spectral fitting package PyQSOFit (\citealp{Guo2018}; \citealp{Shen2019}). We first correct each spectrum to the rest frame and correct for Galactic extinction using the extinction law of \citet{Fitzpatrick1999} and the dust map of \citet{Schlafly2011}. Then, we carry out the host galaxy decomposition using the principal component analysis (PCA; \citealp{Yip2004a}; \citealp{Yip2004b}) implemented in the PyQSOFit code. We apply 5 PCA components for galaxies, which account for $\sim$ 98\% of the galaxy sample, and 20 PCA components for quasars that can reproduce more than 92\% of quasars. Next, the power law, UV/optical Fe II and Balmer component models are applied. After subtracting the best-fit continuum model from each spectrum, we fit the $\rm H\beta$ region (i.e.,  $\rm H\beta \lambda$4863, $\OIII \lambda$5007,4959) and $\rm H\alpha$ region (i.e.,  $\rm H\alpha \lambda$6564.6, $\NII \lambda$6549,6585, $\SII \lambda$6718,6732). The upper limit of the full width at half maximum (FWHM) for the narrow component is set to 900 \kms~to disassemble the narrow and broad components (\citealp{Wang2009}; \citealp{Coffey2019}; \citealp{Wang2019}). Additionally, {we tie the widths and velocities of the narrow components together. For individual emission lines, we characterize the narrow component with a single-Gaussian function and consider an additional broad component when present. The majority of our targets exhibit FWHM ($\rm H\beta$) < 1000 \kms, while in a few targets, $\rm H\beta$ and $\rm H\alpha$ are broad. For these targets, $\rm H\beta$ and $\rm H\alpha$ emission lines require multi-gaussian models (see Figure \ref{fig:example_broad}). The velocity shift up to $\pm$ 3000 \kms~and $\pm$ 1000 \kms~are allowed for the broad and narrow components, respectively. Moreover, the intensity ratios of $\OIII \lambda 5007$/$\OIII \lambda 4959$ and $\NII \lambda 6585$/$\NII \lambda 6549$ are fixed to their theoretical values, i.e., 3 (\citealp{Storey2000}).\\
The distinction of the nongravitational component is relatively straightforward when the line profile is very broad; in contrast, this identification is difficult if this component does not strongly exceed the gravitational component of the bulge. Hence, to clearly distinguish the outflow component in our obscured (U)LIRGs, we define outflow detection when the peak amplitude of the broad outflowing component of the $\OIII \lambda 5007$ emission line is two times larger than the continuum noise. Furthermore, we visually inspect all emission lines to ensure proper fitting.\\
Due to the similar line widths of the narrow and broad components of \OIII~profiles in some sources, it is difficult to distinguish between the outflow and core components. Hence, we calculate the \OIII~velocity shift and dispersion in a similar way to \citet{Woo2016}, who used the first moment (the flux-weighted center) and second moment (velocity dispersion) of the \OIII~profile as follows:}

\begin{equation}\label{equ:mom1}
\lambda_{0} = {\int \lambda f_\lambda d\lambda \over \int f_\lambda d\lambda}.
\end{equation}

\begin{equation}\label{equ:mom2}
\sigma^{2}_{\rm [O\textsc{iii}]} = {\int \lambda^2 f_\lambda d\lambda \over \int f_\lambda d\lambda} - \lambda_0^2, 
\end{equation}
  
where f$_{\lambda}$ is the flux at each wavelength. {We estimate the velocity shift of} \OIII~{with respect to the systemic velocity measured from the pPXF analysis. For a small fraction of sources in which the stellar absorption lines are not characterized reliably (see Section \ref{sec:vel_dispersion}), the peak of the narrow component of H$\rm \beta$ is used to estimate the systemic velocity. To estimate the errors of fitted parameters, we utilize Monte Carlo simulations. For each target, we mock 1000 spectra by randomizing flux with flux uncertainty at each wavelength and fit them. We define the error of each parameter as  $\rm Error\ (X) = 0.5 \times (X_{84.} - X_{6.})$, where $\rm X_{84.}\ and\ X_{6.}$ are the 84th and 6th percentiles of parameter X, respectively. Figures \ref{fig:example_broad} and \ref{fig:example_narrow} illustrate examples of our spectra analyzed with PyQSOFit. Table \ref{tab:properties} summarizes the stellar and ionized gas properties of our (U)LIRGs.}

\begin{figure*} 
\centering 
\includegraphics[width=.95\linewidth]{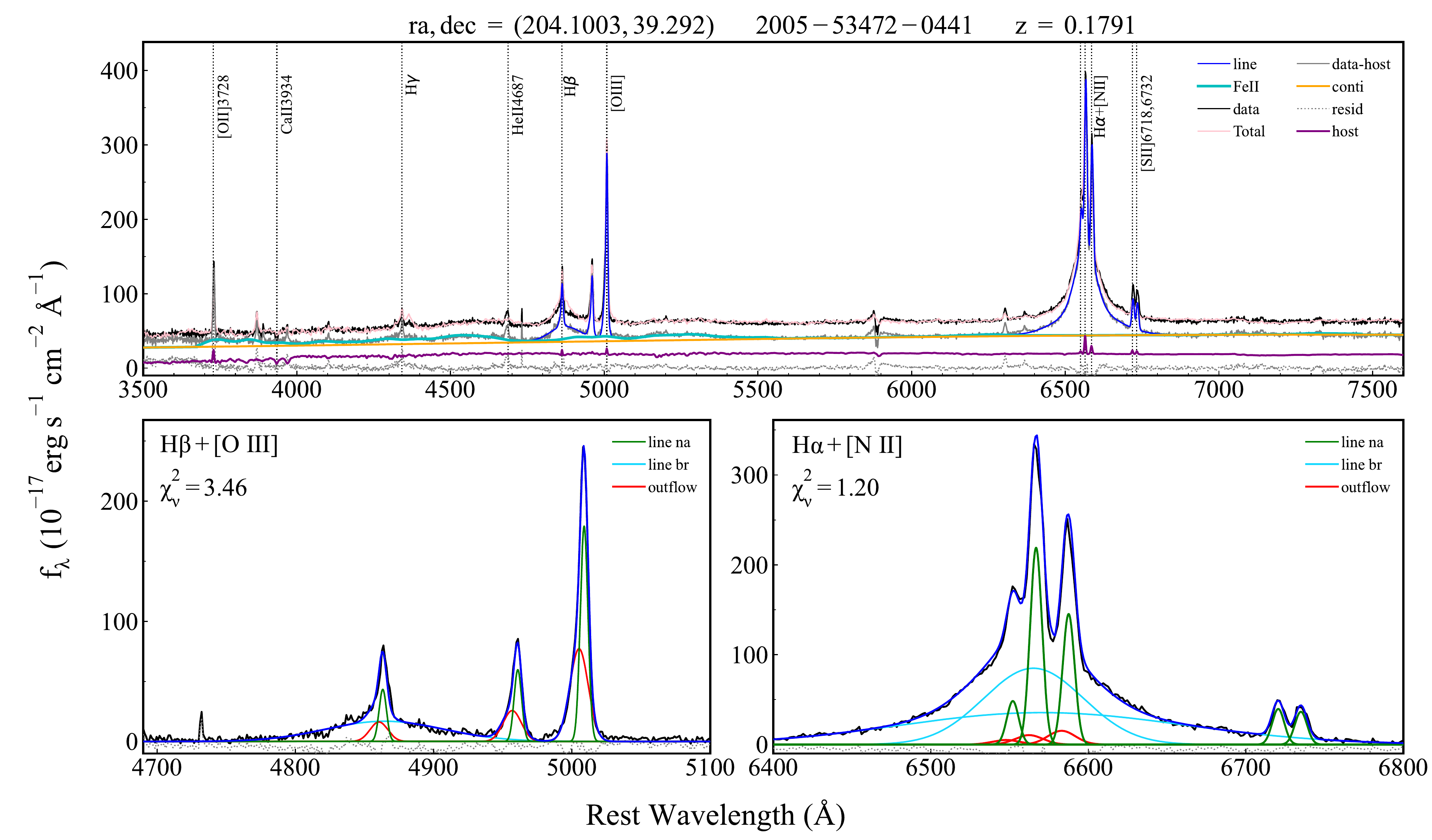}
 \caption{{Example of emission line fitting with PyQSOFit for a target in which H$\rm \beta$ and H$\alpha$ profiles are very broad. Different line components are labeled. The bottom panels are zoomed-in versions of the fitted line complexes.}}
\label{fig:example_broad}
\end{figure*}

\begin{figure*} 
\centering 
\includegraphics[width=.95\linewidth]{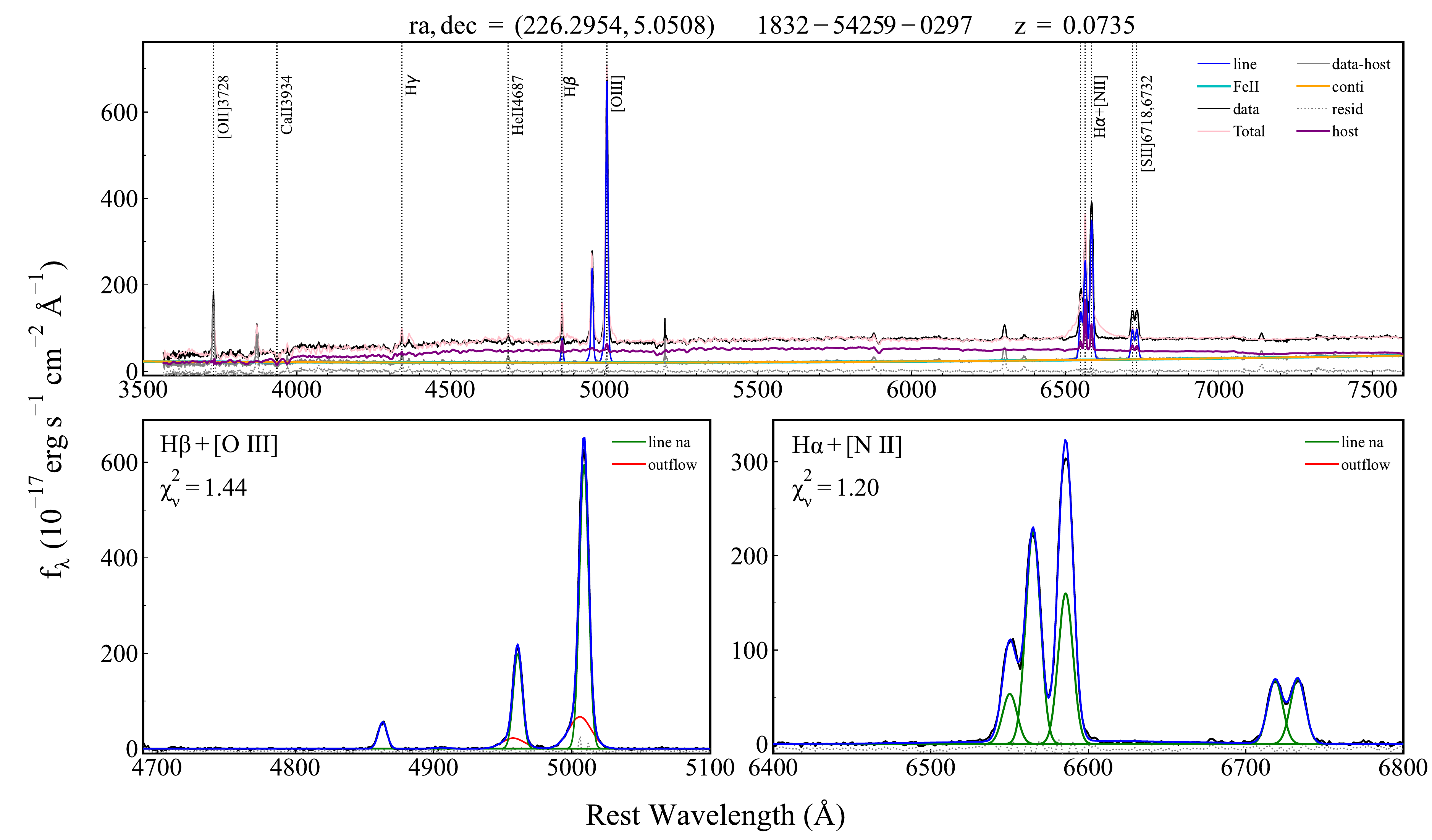}
 \caption{{Same as Figure \ref{fig:example_broad}, but for a target in which H$\rm \beta$ and H$\alpha$ profiles are narrow.}}
\label{fig:example_narrow}
\end{figure*}

Our measurements of the \OIII~velocity shift ($\rm V_{\OIII}$) indicate that the \OIII~profile of AGNs is generally more blueshifted than that of weak/non-AGNs. The value of $\rm |V_{\OIII}|$ reaches up to $\sim$ {810} \kms~and $\sim$ {630} \kms~in AGN and {weak/non-AGN} (U)LIRGs, respectively. 
Additionally, we find that the \OIII~velocity dispersion ($\sigma_{\OIII}$) in AGNs is dramatically higher than that of weak/non-AGNs (KS test $p$-value = $1.5\times10^{-12}$). It implies that AGN activity contributes to broadening line profiles.
In Figure \ref{fig:VVD}, we illustrate the comparison between the \OIII~velocity shift and \OIII~velocity dispersion for our AGN (U)LIRGs. 
Generally, ULIRGs that host AGNs exhibit the most significant the \OIII~velocity shift and \OIII~dispersion. For comparison purposes, we also locate Type 2 AGNs from \citet{Woo2016} (gray circles), a sample of local IR-bright dust-obscured galaxies (DOGs) from  \citet{Toba2017}, hereafter T17 (orange triangles), and a sample of AGN-driven outflows in luminous Type 2 AGNs from \citet{Kang2018} (green triangles) in Figure \ref{fig:VVD}. 
It should be mentioned that some sources in T17 showed larger \OIII~velocity shifts (up to $\sim$ -1500 km s$^{-1}$) and dispersions (up to $\sim$ 1200 km s$^{-1}$) than our sources because their sample mostly included ULIRGs/HyLIRGs\footnote{{HyLIRGs are hyperluminous IR} galaxies with $L_{\rm 8-1000\mu m}\geq 10^{13}$\solarl.} with high AGN contributions, which are expected to have more powerful AGN-driven outflows. Although it is not straightforward to interpret the origin of the velocity shift, we assume that in our AGN (U)LIRGs, AGN emission coupled to star formation activity (e.g., \citealp{Netzer2009}; \citealp{Woo2012}) drives ionized gas outflows detected in the narrow-line region (NLR). Therefore, the highest \OIII~velocity shifts can be associated with sources with powerful AGNs and strong star formation (see Figure \ref{fig:VVD}). In weak/non-AGNs, nuclear starbursts are responsible for photoionizing gas and driving ionized gas outflows (see Section 4.1 of \citealp{Bae2014} for more discussion). 

Furthermore, we reveal that the  \OIII~profile is blueshifted in {$\sim$ 75\%} of our AGN and weak/non-AGN (U)LIRGs. This fraction agrees with the findings of T17 and \citet{Kang2018}, which are $\sim$ 80\% and 74\%, respectively. However, \citet{Woo2016} found a lower fraction of \OIII~blueshifted component in Type 2 AGNs ($\sim$ 50\%). It is probably because the receding component is more obscured in dust-enshrouded (U)LIRGs. According to the models of biconical outflows in NLR and dust extinction, the emission from the side of outflows near the observer is blueshifted and generally less extinct. In contrast, the emission from the receding side is redshifted and preferentially obscured by the inner galactic disk, making this component undetectable or weak at optical bands in some sources (\citealp{Crenshaw2010}).

\begin{figure}  
\includegraphics[width=0.95\linewidth]{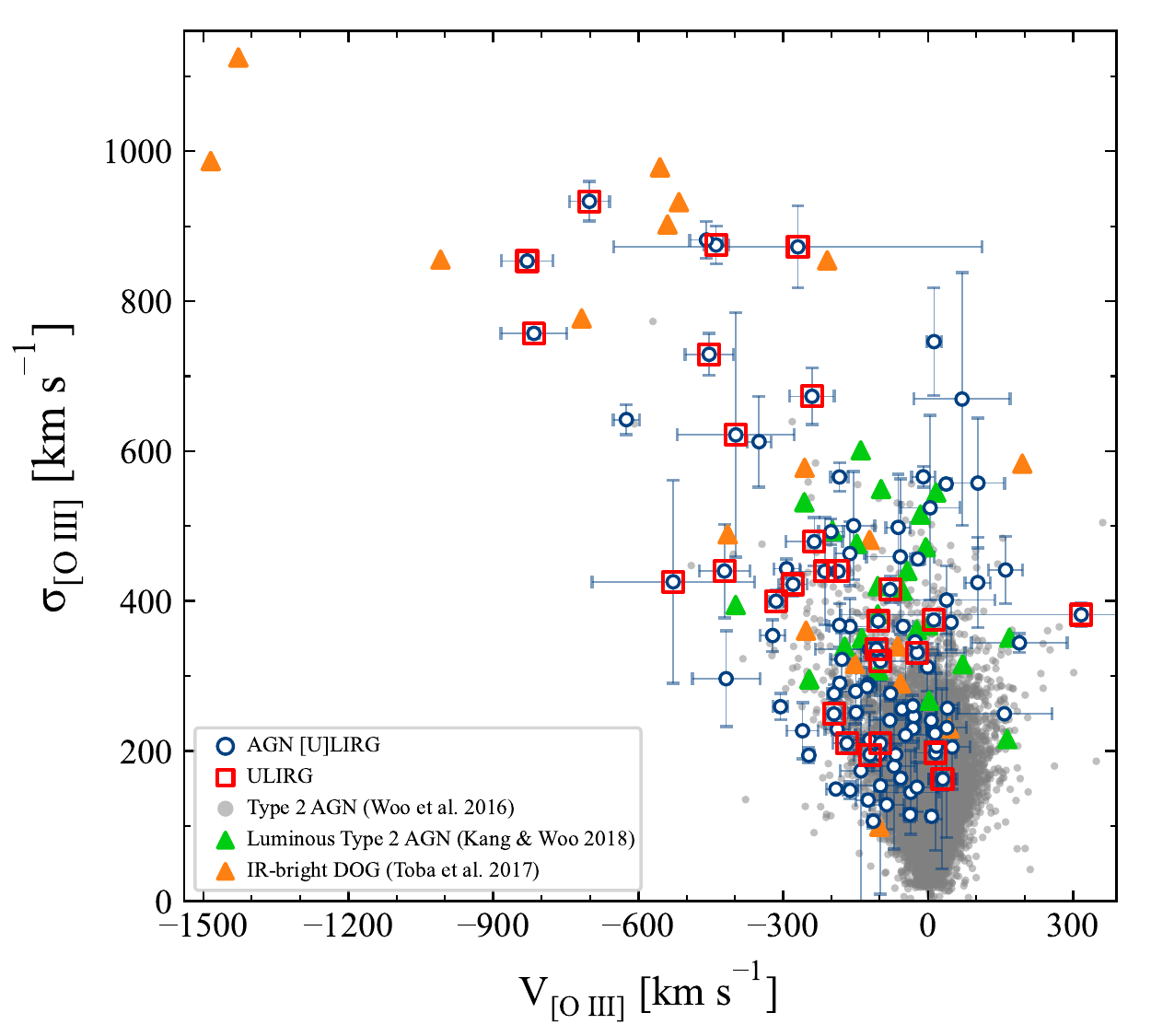}
\caption{Diagram of \OIII~velocity shift versus velocity dispersion. The blue circles and red squares indicate AGN (U)LIRGs and ULIRGs, respectively. The orange and green triangles, and gray dots show DOGs, luminous Type 2 AGNs and Type 2 AGNs from T17, \citet{Kang2018} and \citealp{Woo2016}, respectively.}
\label{fig:VVD}
\end{figure}

\begin{table*}
\scriptsize
\centering
\caption{The stellar and ionized gas outflow properties of our (U)LIRGs.}
\begin{tabular}{|c|c|c|c|c|c|c|c|c|}
\hline
\multirow{2}{*} {ID} & Redshift & log \lumir & $\rm f_{AGN}$ & SFR & \mass & $\sigma_{*}$ &  log $\rm L_{\OIII}^{Corr}$ & $ \sigma_{0}$\\
  &   & [$\rm L_{\odot}$] &  \% & [$\rm M_{\odot}\ yr^{-1}$] & [$\rm 10^{10}\ M_{\odot}$] & [\kms] & [$\rm L_{\odot}$] & [\kms]\\
\hline
5000012  & 0.102  & 11.49 $_{-0.03}^{+0.03}$   & 4.55 $\pm$ 2.24     & 98.21 $\pm$ 46.99    & 10.60 $\pm$ 1.89    & 141.17 $\pm$ 10.95  & 9.54 $\pm$ 0.01   & 400.79 $\pm$ 16.33\\
5000385  & 0.119  & 11.52 $_{-0.01}^{+0.04}$   & 3.51 $\pm$ 2.31     & 83.83 $\pm$ 28.66    & 10.20 $\pm$ 3.69    & 248.44 $\pm$ 24.89  & 7.64 $\pm$ 0.04   & 216.19 $\pm$ 43.71\\
5000566  & 0.122  & 11.53 $_{-0.02}^{+0.04}$   & 1.76 $\pm$ 2.01     & 45.05 $\pm$ 17.74    & 22.90 $\pm$ 3.93    & 155.21 $\pm$ 15.07  & 7.94 $\pm$ 0.02   & 326.21 $\pm$ 39.57\\
5001023  & 0.139  & 11.53 $_{-0.03}^{+0.03}$   & 1.84 $\pm$ 1.34     & 124.90 $\pm$ 29.70  & 8.61 $\pm$ 1.55      & 122.26 $\pm$ 41.42  & 8.13 $\pm$ 0.05   & 199.67 $\pm$ 48.34\\
5002243  & 0.069  & 11.30 $_{-0.03}^{+0.03}$   & 0.73 $\pm$ 0.96     & 20.05 $\pm$ 6.85      & 4.45 $\pm$ 1.21      & 132.89 $\pm$ 17.79  & 6.56 $\pm$ 0.02   & 123.30 $\pm$ 10.68\\
5002288  & 0.125  & 11.78 $_{-0.03}^{+0.03}$   & 7.77 $\pm$ 4.23     & 104.69 $\pm$ 39.16  & 29.00 $\pm$ 12.00  & 193.84 $\pm$ 15.77  & 9.24 $\pm$ 0.02  & 284.58 $\pm$ 21.14\\
5002673  & 0.041  & 11.21 $_{-0.02}^{+0.04}$   & 28.44 $\pm$ 6.47    & 9.31 $\pm$ 1.87        & 8.72 $\pm$ 2.70      & 121.08 $\pm$ 8.60    & 8.65 $\pm$ 0.01   & 287.63 $\pm$ 9.57\\
5003440  & 0.144  & 11.26 $_{-0.05}^{+0.03}$   & 5.65 $\pm$ 4.11     & 54.43 $\pm$ 19.99    & 27.00 $\pm$ 5.13     & 196.44 $\pm$ 15.80  & 8.96 $\pm$ 0.09  & 230.44 $\pm$ 82.28\\
5005561  & 0.081  & 11.47 $_{-0.04}^{+0.02}$   & 1.32 $\pm$ 1.22     & 42.58 $\pm$ 14.57    & 7.77 $\pm$ 2.38       & 184.69 $\pm$ 16.22  & 7.77 $\pm$ 0.02  & 193.98 $\pm$ 16.64\\
5008442  & 0.117  & 11.42 $_{-0.03}^{+0.03}$   & 1.44 $\pm$ 1.66     & 37.36 $\pm$ 25.76    & 13.60 $\pm$ 3.72     & 160.84 $\pm$ 17.08  & 8.32 $\pm$ 0.05  & 351.08 $\pm$ 84.05\\
5009068  & 0.107  & 11.29 $_{-0.04}^{+0.03}$   & 1.16 $\pm$ 0.86     & 55.44 $\pm$ 19.68    & 12.80 $\pm$ 4.29     & 212.71 $\pm$ 22.00  & 6.80 $\pm$ 0.13  & 112.29 $\pm$ 374.43\\
5009143  & 0.094  & 11.29 $_{-0.02}^{+0.04}$   & 1.03 $\pm$ 1.09     & 46.29 $\pm$ 26.50    & 7.59 $\pm$ 1.72       & 101.55 $\pm$ 47.15  & 6.65 $\pm$ 0.12  & 216.65 $\pm$ 72.61\\
5009372  & 0.042  & 11.13 $_{-0.02}^{+0.03}$    & 30.94 $\pm$ 4.42   & 8.68 $\pm$ 3.02        & 4.58 $\pm$ 0.98        & 86.08 $\pm$ 13.62   & 7.80 $\pm$ 0.00  & 241.24 $\pm$ 8.83\\
5010076  & 0.045  & 11.15 $_{-0.03}^{+0.03}$    & 0.34 $\pm$ 0.62     & 24.27 $\pm$ 5.63      & 6.89 $\pm$ 1.42       & 137.31 $\pm$ 9.91    & 8.17 $\pm$ 0.00  & 247.72 $\pm$ 6.04\\
5010364  & 0.058  & 11.14 $_{-0.03}^{+0.03}$    & 1.02 $\pm$ 1.10     & 15.56 $\pm$ 6.04      & 3.58 $\pm$ 1.05       & 194.12 $\pm$ 24.02  & 7.67 $\pm$ 0.01  & 114.35 $\pm$ 13.15\\
\hline
\end{tabular}
{{\bf{Note.}}(1) \akari~ID. (2) Redshift, provided by SDSS. (3) IR luminosity from \citet{Kilerci2018}. (4) AGN fraction (i.e, the IR luminosity contributed from the AGN). (5) Star formation rate. (6) Stellar mass. (7) Stellar velocity dispersion. (8) Extinction-corrected \OIII~luminosity. (9) Sum of the \OIII~velocity dispersion and velocity shift in quadrature. The full Table is available in online version.}
\label{tab:properties}
\end{table*}

\section{Discussions}\label{sec:discussion}
\subsection{Prevalence of Ionized Gas Outflows}
In this subsection, we aim to determine the prevalence of ionized gas outflows in our (U)LIRGs. Based on our definition of detected outflows (see Section \ref{sec:outflows}), we identify that nearly 44\% (435/990) of our sources show the signature of outflows. There is a significant difference in the occurrence of ionized gas outflows between AGN and weak/non-AGN (U)LIRGs as it drops from 72\% (101/141) in AGNs to 39\% (334/849) in weak/non-AGNs. Since our AGN ULIRGs have a detection rate of 85\% (29/34), we investigate their impact on the detection rate in AGNs. Therefore, we exclude them from the AGN subsample and find that the detection rate decreases slightly to 67\% (72/107), which is still higher than that in the weak/non-AGN subsample. It suggests that AGN contribution plays an essential role in driving ionized gas outflows. Using a large sample of Type 2 AGNs, \citet{Woo2016} also found that the outflow detection rate increases from 20\% in low-luminosity AGNs to 90\% in high-luminosity AGNs (i.e., \lumo~> $\rm 10^{42}\ erg\ s^{-1}$). However, \citet{Rojas2020} found little to no dependency of outflow fraction on AGN luminosity ($\rm L_{bol} = 10^{43.5 - 45.5}\ erg\ s^{-1}$). It may be due to their narrow range of AGN luminosities and exclusion of AGNs with weak outflows ($\rm V_{max}$ < 650 \kms). Nonetheless, they found that the incidence of outflows increases with the Eddington ratio. As the more luminous AGNs are manifested by broader \OIII~profiles, the outflow components are more easily detectable. Hence, we expect to see the higher prevalence in luminous sources compared to low-luminosity AGNs. 

Moreover, if we divide our sample into LIRGs and ULIRGs, regardless of AGN contribution, we find that the outflow detection rate {increases} from {43\% (405/952)} in LIRGs to 79\% (30/38) in ULIRGs. As a comparison, \citet{Rodriguez2013} found that nearby ULIRGs show a detection rate of 94\%. Predominantly, the presence of AGNs in ULIRGs is much more frequent than LIRGs (see Figure \ref{fig:wise} and also \citealp{Arribas2014}). While 90\% (34/38) of our ULIRGs host AGNs, less than 11\% (107/952) of LIRGs can be identified as AGNs. Besides, ULIRGs host more intense starbursts than LIRGs (see Figure \ref{fig:sfr_mass}). Hence, the occurrence of more extreme outflows in ULIRGs can be understood.

\subsection{Outflows vs. Host-galaxy Properties}\label{host}
It is difficult to estimate the intrinsic outflow velocity from nonspatially resolved observations due to unknown projection effects and the complex outflow geometry. Nevertheless, we may quantify the strength of outflows and estimate the intrinsic outflow velocity considering appropriate assumptions to study the impact of outflows on their host galaxies. \citet{Bae2016} discussed that assuming a biconical outflow model, the intrinsic outflow velocity can be properly estimated based on a combination of velocity shift and velocity dispersion as: 

\begin{equation}
\label{equation:sigma_0}
\rm \sigma_0 = \sqrt{V_{\OIII}^2 + \sigma_{\OIII}^2}.
\end{equation}
\begin{equation}
\label{equation:V_out}
\rm V_{out} = (2\pm0.5)\,\sigma_{0},
\end{equation}

where $\rm V_{out}$ is the intrinsic outflow velocity (hereafter, outflow velocity). In addition to the increase in the occurrence of outflows with AGN luminosity, theoretical models predict the increase of outflow velocity with AGN luminosity (e.g., \citealp{Costa2014}). Therefore, we examine the correlation between outflow velocity and AGN IR luminosity in our sample (see the left panel in Figure \ref{fig:vout_luminosity}). All correlations throughout this paper are driven using LINMIX\footnote{\url{https://linmix.readthedocs.io/en/latest/src/linmix.html.}} method (\citealp{Kelly2007}). The LINMIX utilizes a hierarchical Bayesian approach to perform the linear regression and can account for measurement errors on both variables in the fit. {We find that the best-fit relation for AGN (U)LIRGs is $\rm V_{\rm out} \propto \lumir(AGN)^{0.23\pm0.04}$,} consistent with T17 who found that $\rm \sigma_{0} \propto \lumir(AGN)^{\sim\ 0.24}$. Note that the correlation in T17 has been driven using a combined sample of IR-bright DOGs and Type 2 AGNs. A similar correlation is also established by \citet{Fiore2017} between the maximum velocity of outflows and a wide range of the AGN bolometric luminosity. This strong relation suggests that the outflows in AGN (U)LIRGs are AGN-driven. In contrast, there is no such correlation in our weak/non-AGN sample ($\rm V_{\rm out} \propto \lumir(AGN)^{0.08\pm0.02}$). It may suggest the weak contribution of AGNs in driving ionized gas outflows in this subsample. 

{We also investigate the correlation between outflow velocity and \OIII~luminosity as a proxy of the AGN bolometric luminosity (\citealp{Heckman2014})}. We correct \OIII~luminosity for Galactic extinction using equation 1 in \citet{Dominguez2013}:

\begin{equation}
\rm L(\lambda)^{\rm corr} = \rm L(\lambda)^{\rm obs}\, 10^{0.4 k(\lambda)\, E(B-V)},
\end{equation}

where $\rm L^{obs}$ is the observed luminosity and k$(\lambda)$ is the reddening at wavelength $\lambda$ and estimated from the \citet{Calzetti2000} curve:

\begin{equation}
\rm k(\lambda) =  2.659 (-2.156 + \frac{1.509}{\lambda} - \frac{0.198}{\lambda^2} + \frac{0.011}{\lambda^3}) + R_{V}.
\end{equation}

$\rm E(B-V)$ is the color excess. {We estimate $\rm E(B-V)$ from Balmer decrement as:}

\begin{equation}
\rm E(B-V) =  \frac{2.5}{k(\lambda_{\rm H\beta}) - k(\lambda_{\rm H\alpha})} \rm log_{10} [\frac{(\rm H\alpha/H\beta)_{\rm obs}} {(\rm H\alpha/H\beta)_{\rm int}}],
\end{equation}

{where $(\rm H\alpha/H\beta)_{obs}\ and\ (H\alpha/H\beta)_{int}$ are the observed and the intrinsic Balmer decrement, respectively.} The typical value of $\rm R_{\rm V}$ = 3.1 is adopted for the ratio of the total-to-selective extinction parameter.

\begin{figure*} 
\centering
\includegraphics[width=.8\linewidth]{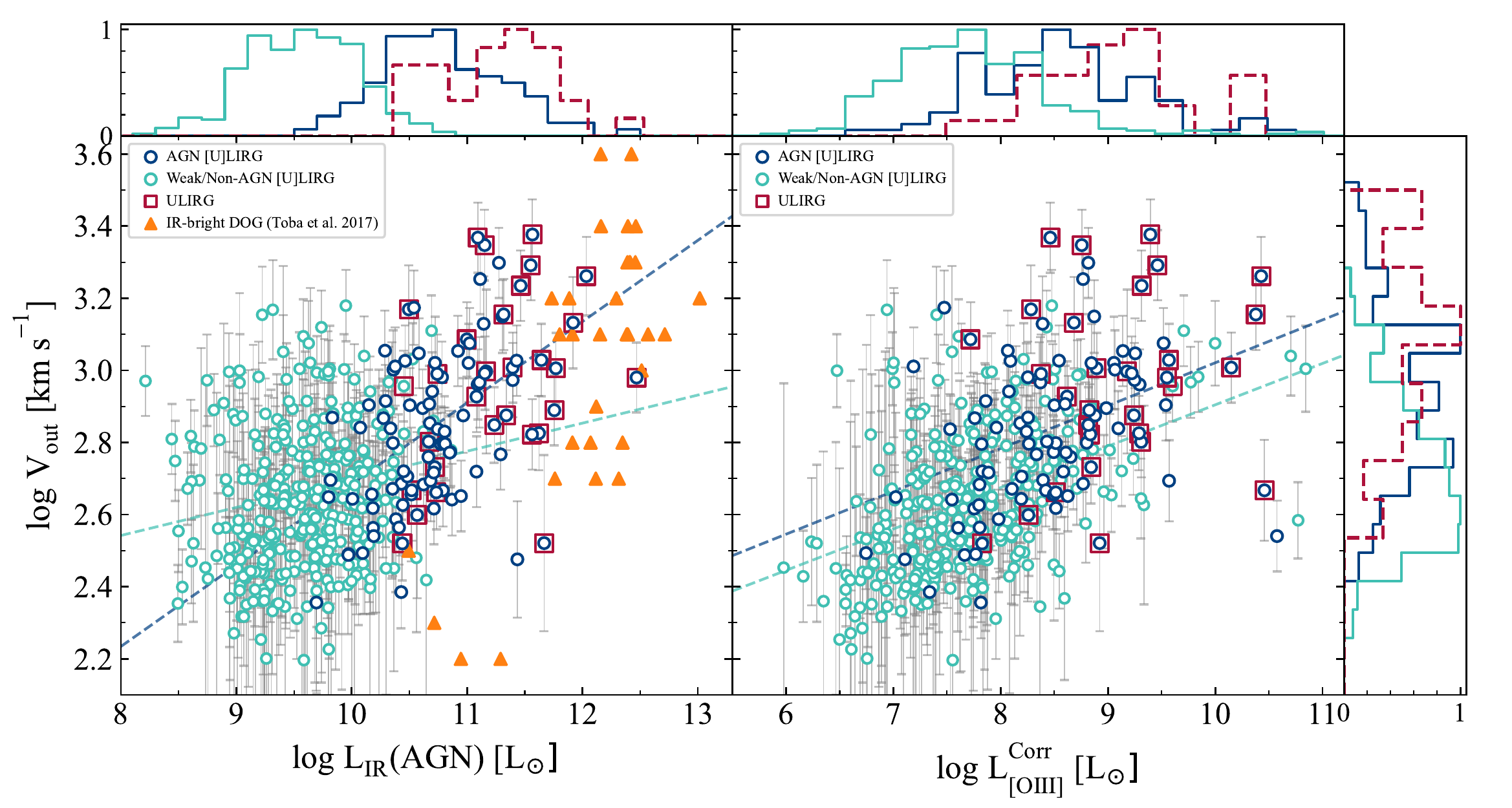}
\caption{Outflow velocity as a function of AGN IR luminosity ($left$) and extinction-corrected \OIII~luminosity ($right$). The colors and symbols are the same as in Figure \ref{fig:sfr_mass}. The orange triangles present IR-bright DOGs from T17. The blue and green dashed lines show the best-fit for AGN and weak/non-AGN (U)LIRGs, respectively.}
\label{fig:vout_luminosity}
\end{figure*}

The right panel of Figure \ref{fig:vout_luminosity} presents the outflow velocity as a function of extinction-corrected \OIII~luminosity. We find that outflows with the highest velocities can be driven by the most \OIII-luminous galaxies (see also \citealp{Leung2019}; \citealp{Rojas2020}). 
For AGN (U)LIRGs, {the best-fit correlation between the extinction-corrected \OIII~luminosity and $\rm V_{out}$ is obtained as} {$\rm V_{out} \propto \lumo^{0.13\pm0.03}$. This relation is flatter than the $\rm V_{out}$--\lumir(AGN) correlation found above. The correlation based on our AGN subsample is also not as steep as that found by \citet{Leung2019} where the outflow velocity is proportional to $\lumo^{0.27\pm0.09}$ for a sample of AGNs (including both quiescent and star-forming galaxies) at 1.4 < $z$ < 3.8. Since our sample is significantly above the main sequence, one explanation may be the more severe contamination of star formation in \OIII~luminosity. In contrast to the AGNs, {we find that outflow velocity in weak/non-AGNs correlates with the extinction-corrected \OIII~luminosity as $\rm V_{out} \propto \lumo^{0.12\pm0.01}$ which is much stronger than the correlation between outflow velocity and IR AGN luminosity, likely due to the contribution of both AGN activity (however not very significant) and star formation in these galaxies.

Examining of the relationship between the ionized gas outflow properties and SFR as well as specific SFR (sSFR=SFR/\mass) may provide some information about the feedback mechanisms. Here, we first examine how the ionized gas outflow velocity of our AGN (U)LIRGs might be connected to the host galaxy velocity dispersion, and then we explore the correlation between outflow velocity and the SFR derived by SED-fitting. Using a sample of 68 Seyfert galaxies, \citet{Nelson1996} systematically compared the \OIII~velocity dispersion and stellar velocity dispersion for first time. They found a good correlation between \OIII~kinematics and stellar velocity dispersion. It suggests that the bulge gravitational potential governs the \OIII~kinematics. However, this correlation was found with a considerably large scatter, which is indicative of an additional kinematic component (nongravitational component). In particular, \citet{Greene2005} showed that \OIII~velocity dispersion is well correlated with stellar velocity dispersion only after removing the broad component. It implies that when a broad component is present in \OIII~profile, the velocity dispersion of ionized gas is not a good representative of the host-galaxy velocity dispersion. These results were later confirmed using much larger samples of AGNs (e.g., \citealp{Woo2016}). In Figure \ref{fig:sigma_sigma} we present the distribution of stellar velocity dispersion as a function of \OIII~velocity dispersion of our AGN sample in comparison with Type 2 AGNs from \citet{Woo2016}. The black dotted line shows the one-to-one relation between $\sigma_{[\rm O\ III]}$ and $\sigma_{*}$, while, the blue dashed line indicates the average ratio of $\sigma_{[\rm O\ III]}/\sigma_{*} \sim 1.4$ for Type 2 AGNs with double Gaussian \OIII~profile (\citealp{Woo2016}). The ratio of $\sigma_{[\rm O\ III]}/\sigma_{*}$ for our AGN (U)LIRGs, on average, is $\sim 1.8-1.9$, larger than the ratio for Type 2 AGNs in \citet{Woo2016}. It suggests that our objects exhibit more extreme ionized gas outflows than Type 2 AGNs (see also Figure \ref{fig:VVD}); thus the contribution of the host-galaxy gravitational potential is less critical in driving \OIII~kinematics in our sample. \citet{Woo2016} suggested that the effect of the bulge potential can be corrected through normalizing \OIII~velocity width by the stellar velocity dispersion.} They utilized the value of $\sigma_{0}/\sigma_{*}$ to classify outflows into the strong (log $\sigma_{0}/\sigma_{*}$ \> 0.3), intermediate (0 \< log $\sigma_{0}/\sigma_{*}$ \< 0.3), and weak (log $\sigma_{0}/\sigma_{*}$ \< 0) ones. Therefore, we also compare our estimates of outflow velocities with the normalized quantity of $\sigma_{0}/\sigma_{*}$ that serves as another tracer of outflow strength. The values {of} $\sigma_{0}/\sigma_{*}$ are color-coded for our AGN (U)LIRGs in Figure \ref{fig:vout_sfr_agn}. 

\begin{figure} 
\centering
\includegraphics[width=0.95\linewidth]{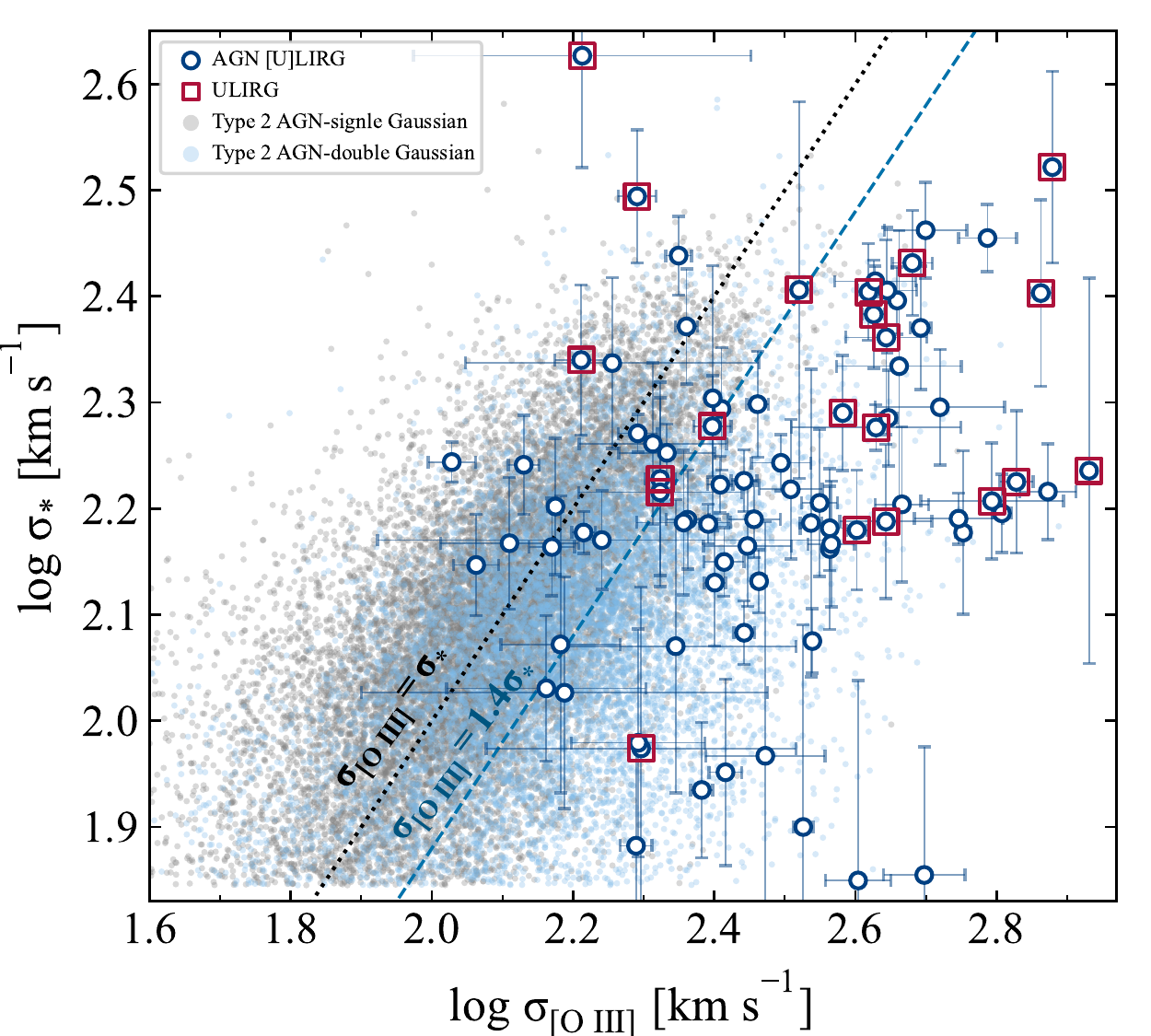}
\caption{Stellar velocity dispersion versus \OIII~velocity dispersion. The colors and symbols are the same as in Figure \ref{fig:VVD}. The gray and blue circles exhibit Type 2 AGNs whose \OIII~profiles have been fitted with single or double Gaussian model (\citealp{Woo2016}). The black dotted line indicates one-to-one correlation. The blue dashed line presents the average ratio of $\sigma_{\OIII}/\sigma_{*} \sim 1.4$ for Type 2 AGNs with double Gaussian \OIII~components.}
\label{fig:sigma_sigma}
\end{figure}
 
As the left panel of Figure \ref{fig:vout_sfr_agn} shows, there is a positive relationship between SFR and outflow velocity in our AGN sample as $\rm V_{out} \propto SFR^{0.13 \pm 0.04}$. In this context, \citet{Leung2019} found that the ionized gas outflow incidence is independent of SFR (obtained by SED-fitting) for their AGN sample. As discussed in \citet{Leung2017}, this lack of strong relationship between outflow properties and SFR should not be translated as the lack of AGN feedback. SFRs estimated from different approaches reflect star-formation activities on different timescales, and the SFR derived from the SED-fitting method is averaged over the last 10$^{8}$ years (\citealp{Kennicutt1998}), while the currently observed outflows are expected to have shorter timescales. For instance, \citet{Leung2019} estimated that their ionized gas outflows observed with the MOSFIRE Deep Evolution Field (MOSDEF) survey at $z\sim$ 2 have dynamical timescales in orders of 10$^{5-7}$ years. Therefore, the observed outflows are not expected to suppress or enhance SFR. \citet{Woo2017} also suggested a delayed feedback scenario using a large sample of local AGNs and argued that it takes the order of a dynamical time for outflows to influence the ISM. Besides, most recently, \citet{Burtscher2021} found that AGNs and star-forming galaxies have similar nuclear ($\leq$ 50 pc) star-formation histories, which can be interpreted as another supporting evidence of the delayed AGN feedback. 

Furthermore, we test the SFR-$\rm V_{out}$ relation with another SFR proxy independent of our SED-fitting procedure. We use the far-IR photometric measurements of $AKARI$/FIS at 90 $\mu$m where the contamination of AGN is insignificant. All our AGNs have high-quality detection (i.e., quality flag $FQUAL$ (90 $\mu$m) = 3). We employ the equation by \citet{Kennicutt1998} to estimate SFR as

\begin{equation}
\rm SFR_{90\mu m}\: [\solarm\: yr^{-1}] = 4.5 \times10^{44}\: \times L_{FIR}\: [\solarl],
\end{equation}

Note that $\rm L_{FIR}$ in the above equation represents the total luminosity integrated over the far-IR, while we use the monochromatic luminosity at 90 $\mu$m (see also \citealp{Woo2020}). The difference between the SED-based SFR (SFR$_{SED}$) and FIR-based SFR (SFR$_{90\mu m}$) on average is $\sim$ 0.13 dex. Therefore, they are comparable. The bottom panel of Figure \ref{fig:vout_sfr_agn} presents the SFR derived from the FIR luminosity (SFR$_{90\mu m}$) as a function of outflow velocity. It is clear that FIR-based SFR makes a correlation with outflow velocity similar to SED-based SFR as $\rm V_{out} \propto SFR_{90\mu m}^{0.15 \pm 0.05}$. However, the positive correlation between SFR and outflow velocity in our sample is established with large scatter, it may imply that AGNs with faster outflows are hosted by galaxies with higher SFRs}. This result is consistent with the recent findings of \citet{Woo2020} for a large sample of local Type 1 and Type 2 AGNs with SFR obtained either from the monochromatic luminosity at 90 (100) $\mu$m from AKARI/FIS (Hershel/PACS) or based on the artificial neural network (ANN).\

 At higher redshifts, \citet{Wylezalek2016} (hereafter WZ16) also found a positive correlation of SFR (estimated from $Herschel$ FIR data and utilizing the method presented in \citealp{Symeonidis2008}) with outflow velocity (90\% velocity width of the \OIII$\lambda$ 5007{\AA} line) in the highest bin of AGN IR luminosity (log\,$\rm \nu L_{\nu}[12 \mu m]~\>~44.9$ erg s$^{-1}$) in a sample of Type 2 AGNs at $z\sim$ 0.1--1. There are different interpretations of this positive correlation. \citet{Woo2017} discussed that the positive correlation between outflow velocity and SFR suggests a lack of instantaneous suppression of star formation activity. They considered an evolutionary sequence as the consequence of either the delayed AGN feedback scenario (i.e., outflows suppress SFR after the dynamical timescales) or the depletion of gas supply. Alternatively, WZ16 interpreted this positive correlation by considering the role of the host-galaxy gas content in accelerating outflows. In other words, in AGNs with high SFRs (which reflect high intrinsic gas contents in host galaxies), the radiation and outflows driven by the AGN couple with the ISM more effectively and faster outflows are produced.

Since the sSFR can constrain the relative effect of outflows and feedback mechanism on the host-galaxy, it has been widely advocated instead of SFR.  The detection of the negative correlation between sSFR and outflow strength may imply a negative AGN feedback scenario by suppressing star formation, i.e., preventing gas from cooling and/or sweeping away reservoirs of gas from the host-galaxy (see \citealp{Harrison2018} for a review). On the other hand, when sSFR increases with outflow velocity, a positive AGN feedback mechanism might be construed. Some recent observations provided direct evidence of the positive AGN feedback through the detection of star formation inside the powerful outflows (\citealp{Maiolino2017}; \citealp{Bao2021}). Therefore, in this part, we investigate the relation between outflow velocity and sSFR in our AGN sample. {As Figure \ref{fig:vout_ssfr_agn} shows, we find no correlation between sSFR and outflow velocity when we consider the entire sample. Following WZ16, we also test the relationship between sSFR and $\rm V_{\rm out}$ for our AGN (U)LIRGs in two different bins of SFR, i.e., SFR < 100 \solarm~yr$^{-1}$ and SFR > 100 \solarm~yr$^{-1}$. We find that outflow velocity of AGN (U)LIRGs with SFR > 100 \solarm~yr$^{-1}$ decreases with sSFR as  $\rm V_{out} \propto sSFR_{SED}^{-0.44 \pm 0.14}$. Our findings are in agreement with results reported by WZ16 who found a strong anti-correlation between  sSFR and $\rm V_{\rm out}$ for gas-rich galaxies with SFR > 100 \solarm~yr$^{-1}$. It may be indicative of suppressed SFR by AGN feedback.}

In this framework, \citet{Woo2017, Woo2020} found that AGNs with strong ionized gas outflows have comparable sSFR to that of the main sequence galaxies, while AGNs with weak/no outflows have much lower sSFR, implying no instantaneous impact of outflows on SFR (i.e., delayed AGN feedback).

\begin{figure*}  
\centering
\includegraphics[width=.7\linewidth]{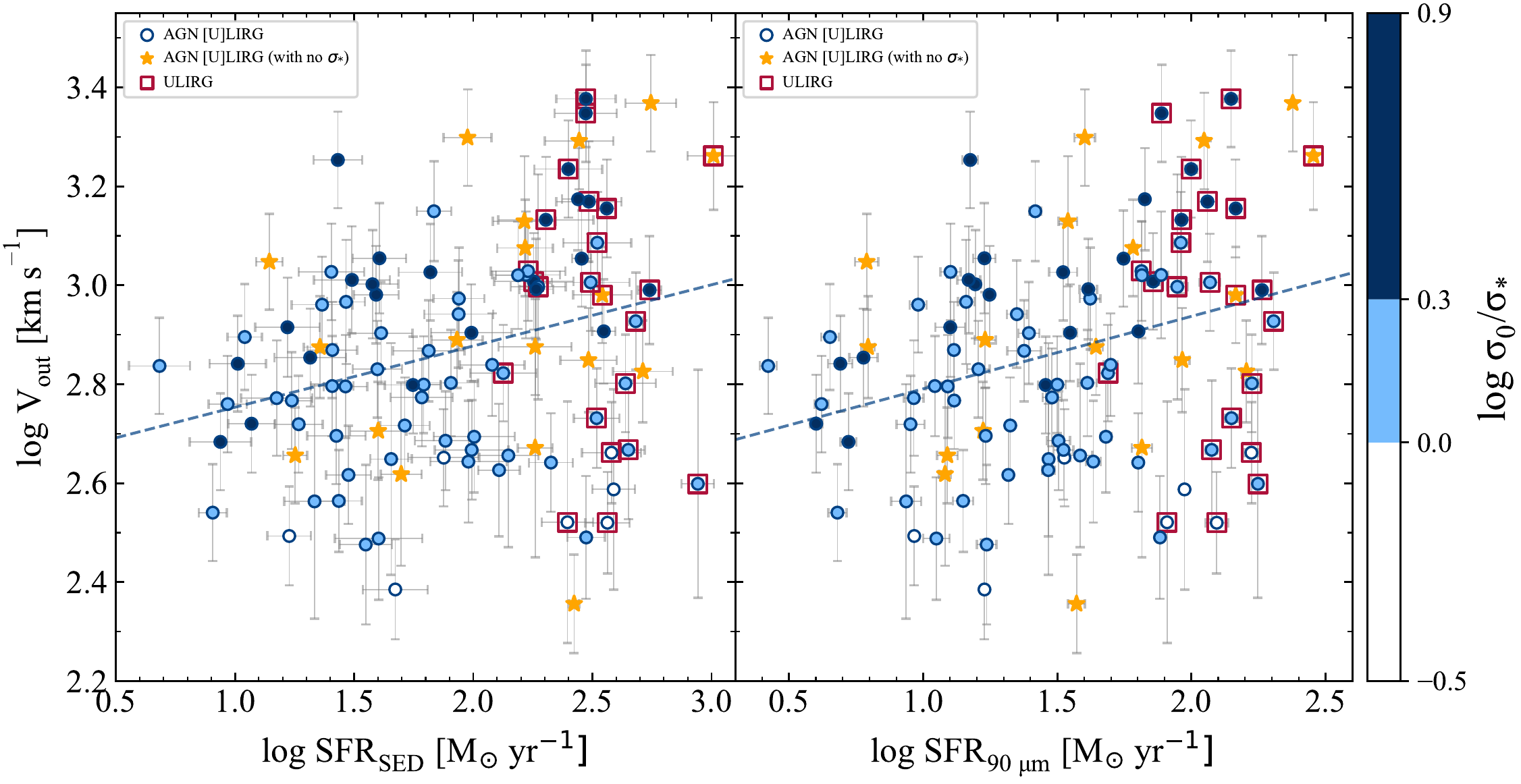}
\caption{Outflow velocity as a function of SFRs estimated from SED-fitting ($left$) and $90\ \mu$m luminosity ($right$) for AGN (U)LIRGs with color-coded $\sigma_{0}/\sigma_{*}$ values. The blue lines represent the regression for our AGN sample. The yellow stars show targets with no reliable stellar velocity dispersion estimates.}
\label{fig:vout_sfr_agn}
\end{figure*}

\begin{figure}  
\includegraphics[width=7.8cm]{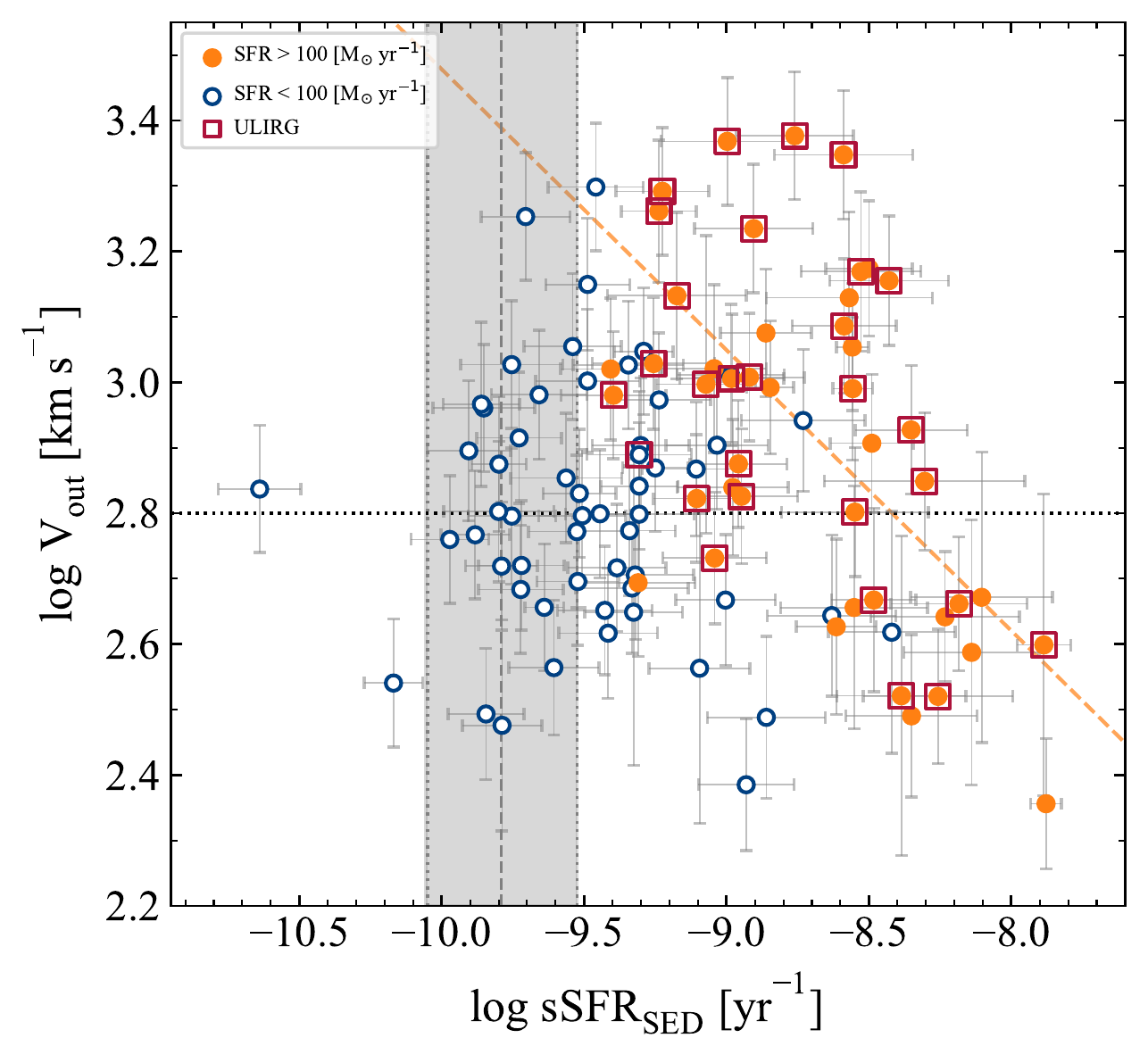}
\caption{Outflow velocity as a function of sSFR for AGN (U)LIRGs. Sources with SFR \> 100 \solarm~yr$^{-1}$ and SFR \< 100 \solarm~yr$^{-1}$ are presented by filled orange and open blue circles, respectively. The red squares indicate ULIRGs. The shaded region exhibits the main sequence of star-forming galaxies from \citet{Woo2020}. The orange dashed line presents the best-fit correlation for AGNs with SFR > 100 \solarm~yr$^{-1}$.}
\label{fig:vout_ssfr_agn}
\end{figure}

In the case of weak/non-AGN (U)LIRGs, our sample covers a too narrow range of SFR (only $\sim$ one order of magnitude) and we do not find significant correlation between SFR and outflow velocity. Therefore, we trace stellar feedback by considering the dependency of outflow velocity on \mass~which has a slightly wider range. Figure \ref{fig:vout_mass_non} shows our sample in combination with the sample of \citet{Chisholm2015}. We also consider a sample of non-AGN (U)LIRGs at $z\sim 0$ from \citet{Arribas2014} who detected ionized gas outflows from H$\alpha$ emission line. For comparison, we assume that their dynamical mass is similar to the \mass. We find a positive correlation between \mass~and outflow velocity in weak/non-AGN (U)LIRGs. Using the LINMIX method, {the best-fit is $\rm V_{out} \varpropto \mass^{0.15~\pm~0.05}$ in our sample (the green dashed line in Figure \ref{fig:vout_mass_non}).} It is in agreement with the scaling relation found by \citet{Chisholm2015} as $\rm V_{out} \varpropto \mass^{\sim 0.12}$ for a sample of nearby star-forming galaxies with ionized gas outflows detected from [Si II] kinematics.
Generally speaking, it is expected that SFR and \mass~correlates tightly with the velocity of starburst-driven outflows, when the more powerful supernova (SN) explosions and more massive stars deposit faster-moving flows of gas into the host-galaxy. However, due to {the} dependency of SFR and \mass~on properties of dust in galaxy, the emergence of a clear picture is difficult. 

\begin{figure} 
\includegraphics[width=8cm]{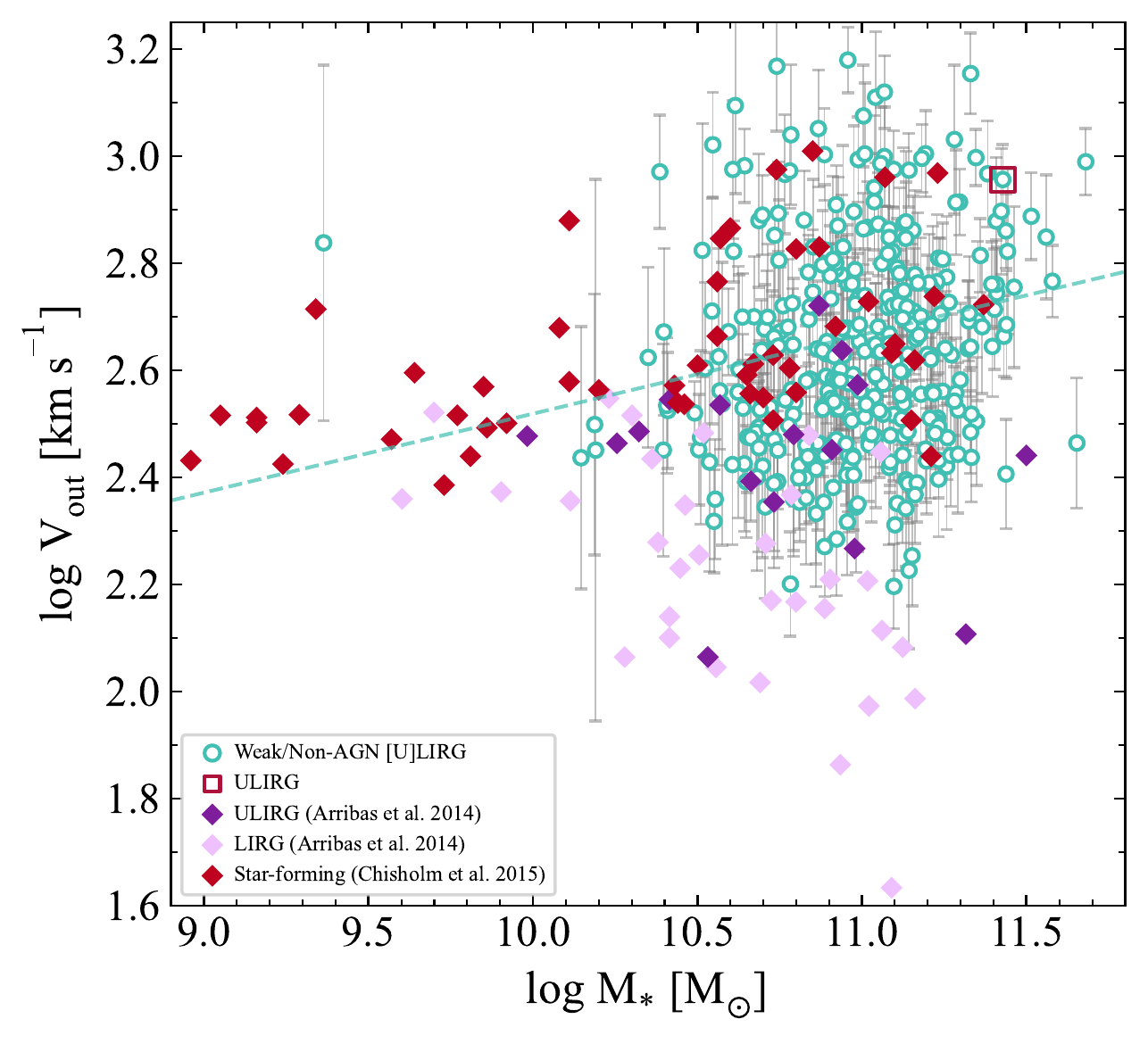}
\caption{Outflow velocity as a function of stellar mass for weak/non-AGN (U)LIRGs. We show our sources with green circles. The sample from \citet{Arribas2014} {is} shown with dark and light purple squares corresponding to ULIRGs and LIRGs, respectively. The red diamonds indicate star-forming galaxies from \citet{Chisholm2015}.}
\label{fig:vout_mass_non}
\end{figure}

\subsection{Mass- and Energy-Outflow Rates}
Outflows impact their host galaxies through the transition of mass and energy. In this subsection, we discuss how the mass-outflow rate ($\rm \dot M_{out}$) and energy rate ($\rm \dot E_{out}$) change with AGN luminosity in our AGN (U)LIRGs, though the precise calculation of these quantities requires knowing the geometry and kinematics of outflowing gas. The density of outflowing gas is commonly subjected to the large uncertainties in measurements of mass and energy rate of outflows. Typically, the electron density is estimated based on the [SII] doublet which is difficult to be detected in outflows. Most recently, \citet{Fluetsch2020} presented the properties of spatially resolved multiphase outflows in 31 local (U)LIRGs and they concluded that the average density of ionized outflowing gas in their (U)LIRGs is 485 $\rm cm^{-3}$. This value is in agreement with findings of Arribas et al. (2014) for local (U)LIRGs ($\rm n_{e} = 459 \pm~66\ cm^{-3}$). However, to compare our results with those of T17, following them, we take electron density to be $\rm n_{e} = 100\ cm^{-3}$ to estimate the mass of the ionized gas in the outflows. This value has also been adopted by other similar works (e.g., \citealp{Brusa2015}). Following \citet{Nesvadba2011}, we estimate the mass of ionized gas from H$\rm \beta$ luminosity as:

\begin{equation}
\rm \frac{M_{out}} {2.82\times 10^9 \solarm} = \left(\frac{L_{{\rm H}\beta}}{10^{43}\,{\rm erg}\,{\rm s}^{-1}} \right) \left(\frac{n_{\rm e}}{100\, {\rm cm}^{-3}} \right)^{-1},
\label{equ:mass_hbeta}
\end{equation}

{where $\rm L_{H\beta}$ is H$\rm \beta$ luminosity (in units of erg s$^{-1}$).} The mass and energy-outflow rates can be expressed as:

\begin{equation}
\rm \dot M_{\rm out} = B \left(\frac{M_{\rm out} V_{\rm out}}{R_{\rm out}}\right),
\label{equ:mass-rate}
\end{equation}
\begin{equation}
\rm \dot{E}_{\rm out} =  \frac{1}{2} \dot{M}_{\rm out} V_{\rm out}^2,\\
\end{equation}

where the factor of B in Equation (\ref{equ:mass-rate}) is a geometry-dependent constant in the range of 1--3 (\citealp{Harrison2018}).  Assuming a spherical sector, we choose $B$ = 3 (see \citealp{Alfonso2017} for details). The parameter of $\rm R_{out}$ is the outflow {radius}. \citet{Bae2017} estimated $\rm R_{out}$ based on the 1D {distribution} of the \OIII~velocity dispersion as a function of radial distance from the center and found that $\rm R_{out}$ is about two times larger than the size of {narrow-line} region ($\rm R_{NLR}$). Hence, we assume $\rm R_{out} = 2 R_{NLR}$ and the size of NLR can be estimated from the extinction-uncorrected \OIII~luminosity as follows (see Section 4.3 in \citealp{Bae2017}):
\begin{equation} 
\rm \log \,R_{NLR} = (0.41 \pm 0.02) \log \,L_{[O III]} - (14.00 \pm 0.77).
\end{equation}

Figure \ref{fig:m_dot_E_dot} shows the mass and energy outflow rates as a function of AGN IR luminosity for our AGN (U)LIRGs. Clearly, the mass outflow rate increases with AGN IR luminosity. About 40\% of our AGN (U)LIRGs have $\rm \dot M_{out}$\> 10 \solarm~yr$^{-1}$. In comparison with the DOG sample of T17, our sources, which are fainter, have systematically smaller values of $\rm \dot M_{out}$. We also find that the kinetic power of the ionized gas outflows driven by our AGN (U)LIRGs has a range of 0.1\% - 30\% of AGN IR luminosity, mostly less than 1\%. These fractions are comparable with {the} kinetic power of DOGs and are higher than those found in low-luminosity AGNs (\citealp{Rojas2020}). According to Figure \ref{fig:m_dot_E_dot}, the ionized gas outflows in high-luminosity sources tend to drive out a larger amount of outflowing gas and higher injected kinetic energy into the host-galaxy.

\begin{figure} 
\centering 
\includegraphics[width=0.9\linewidth]{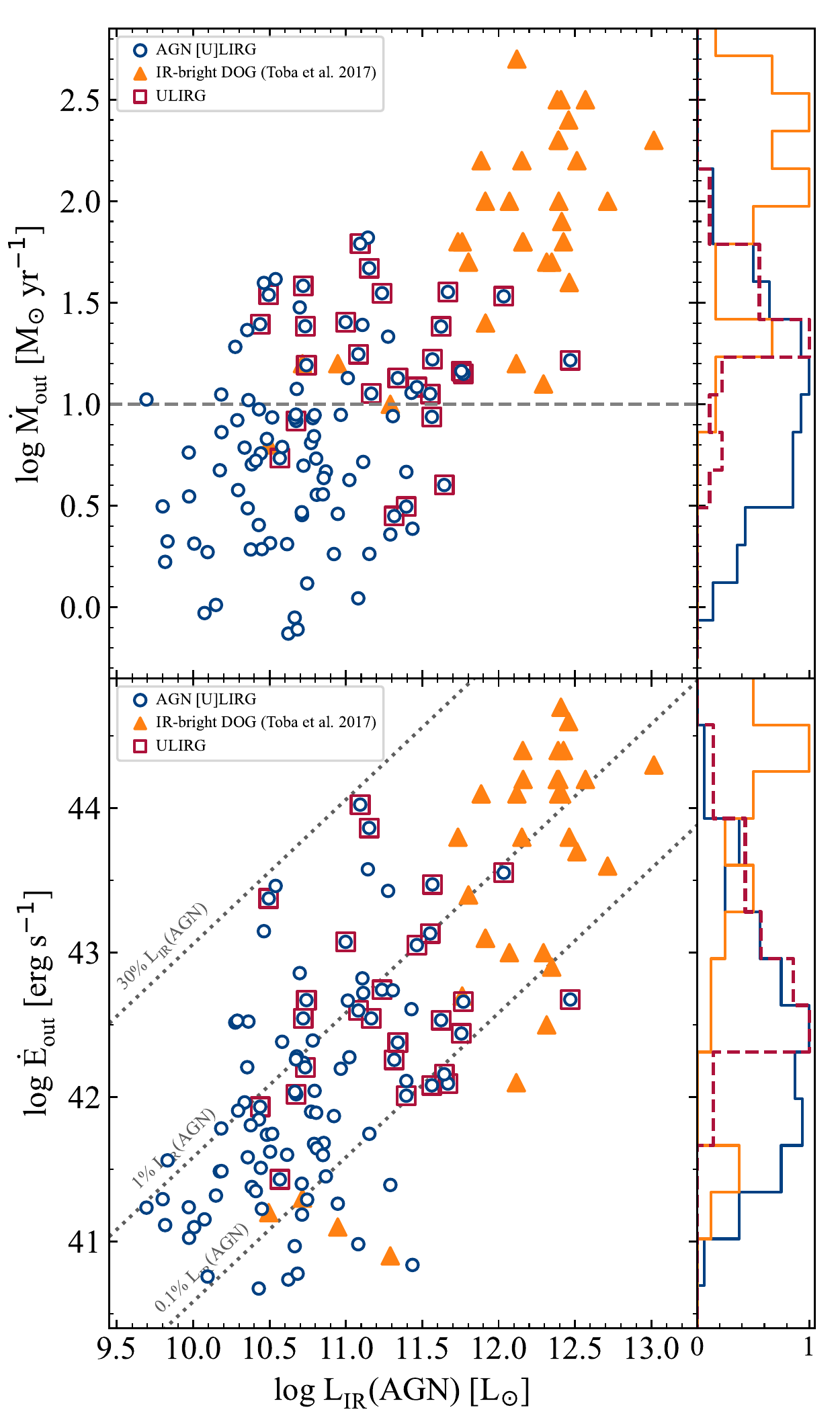}
\caption{Mass ($top$) and energy ($bottom$) outflow {rates} as a function of AGN IR luminosity. The filled orange triangles show DOGs in T17. Other colors are identical to {that in} Fig. \ref{fig:sfr_mass}. The gray dashed line in the $top$ panel indicates $\rm \dot M_{\rm out}$=10 \solarm~yr$^{-1}$.}
\label{fig:m_dot_E_dot}
\end{figure}

\section{CONCLUSIONS}\label{sec:results}
In this study, we explore the stellar kinematics and ionized gas outflows for a large sample of 1106 (U)LIRGs (mostly LIRGs) at 0.02 \< $z$ \< 0.3. Using {the} \akari, \wise~and SDSS data, we fit SEDs of our sources to estimate their physical properties such as stellar mass, SFR and contribution of AGNs in IR emission {(\afrac)}. We divide our sample into AGN and weak/non-AGN (U)LIRGs based on either {the} \wise~color criterion from \citet{Stern2012} or \afrac. We fit \OIII, $\rm H\beta$ and $\rm H\alpha$ lines to characterise ionized gas outfows in our sample. The main conclusions are summarized as follows:

\begin{enumerate}

\item
Almost all our sources reside above the main sequence. It indicates that our (U)LIRGs are very active in star formation. The highest SFRs and stellar masses are associated with ULIRGs. We find no significant difference in stellar-mass distribution between AGN and {weak/non-AGN (U)LIRGs;} however, AGN (U)LIRGs {tend} to have slightly higher {SFRs}.

\item
{Among the 990 (U)LIRGs that have reliable SED-fitting results, there are 435 sources (44\%) that show signatures of ionized gas outflows. The detection rate of outflows increases from 39\% (334/849) in weak/non-AGNs to 72\% (101/141) in AGNs. Furthermore, the outflow detection rate increases from 43\% (405/952) in LIRGs to 79\% (30/38) in ULIRGs.} 

\item
{While} the AGN (U)LIRGs include the fastest outflows with outflow velocity ($\rm V_{out}$) up to $\sim$ 2300 km s$^{-1}$, the maximum {outflow velocity in weak/non-AGN} (U)LIRGs is $\sim$ 1500 km s$^{-1}$, indicative of AGN contribution in driving powerful outflows. 

\item
We find a positive correlation between outflow velocity and SFR (derived through either SED-fitting or 90 $\mu$m luminosity). It implies that AGNs with faster outflows are hosted by galaxies with higher SFRs. However, we find that in AGNs with highest SFRs there is a negative correlation between sSFR and outflow strength. It may suggest the negative feedback scenario in these AGN (U)LIRGs.

\item
In weak/non-AGN (U)LIRGs, the outflow velocity correlates with \mass~as $\rm V_{out} \varpropto \mass^{0.15 \pm 0.05}$ which is in agreement with other studies.

\item
The mass-outflow rate and kinetic power of outflows increase with AGN IR luminosity ($\rm L_{IR}(AGN)$). According to our estimates, AGN ULIRGs which include the most extreme sources have higher values of $\rm \dot M_{out}$ and $\rm \dot E_{out}$ than low-luminosity AGNs. The kinetic power of outflows in our sample has a range of $\sim$ 0.1-30\% {of} $\rm L_{IR}(AGN)$.
\\

\end{enumerate}
We note that in this work we cannot rule out the contamination of starbursts in AGN (U)LIRGs, or vice versa, the contamination of AGNs in {weak/non-AGN} (U)LIRGs. To {better} determine the origin of outflows and the contribution of each component, spatially resolved and higher-spectral-resolution observations would be desired.
\\

\noindent {\bf Acknowledgements}
We thank the anonymous referee for detailed comments, which were useful in improving our manuscript. We also thank Prof. Jong-Hak Woo for insightful discussions and for generously sharing the Type 2 AGN catalog used in this study. A.A acknowledges support from China Scholarship Council for the Ph.D. Program (No. 2017SLJ021244). A.A acknowledges support from the Basic Science Research Program through the National Research Foundation of the Korean Government (grant No. NRF- 2021R1A2C3008486). A.A., Y.Q.X. and H.A.N.L. acknowledge support from the National Natural Science Foundation of China (NSFC-12025303, 11890693, {12003031}), National Key R \& D Program of China No. 2022YFF0503401, the K.C. Wong Education Foundation, {and the science research grants from the China Manned Space Project with NO. CMS-CSST-2021-A06.}  


\appendix

Figure \ref{fig:mock} shows the comparison between the true values provided by the code and the values estimated with the mock catalog.
\renewcommand{\thefigure}{A.\arabic{figure}}
\setcounter{figure}{0}

\begin{figure}[hbt!]
\includegraphics[width=\linewidth]{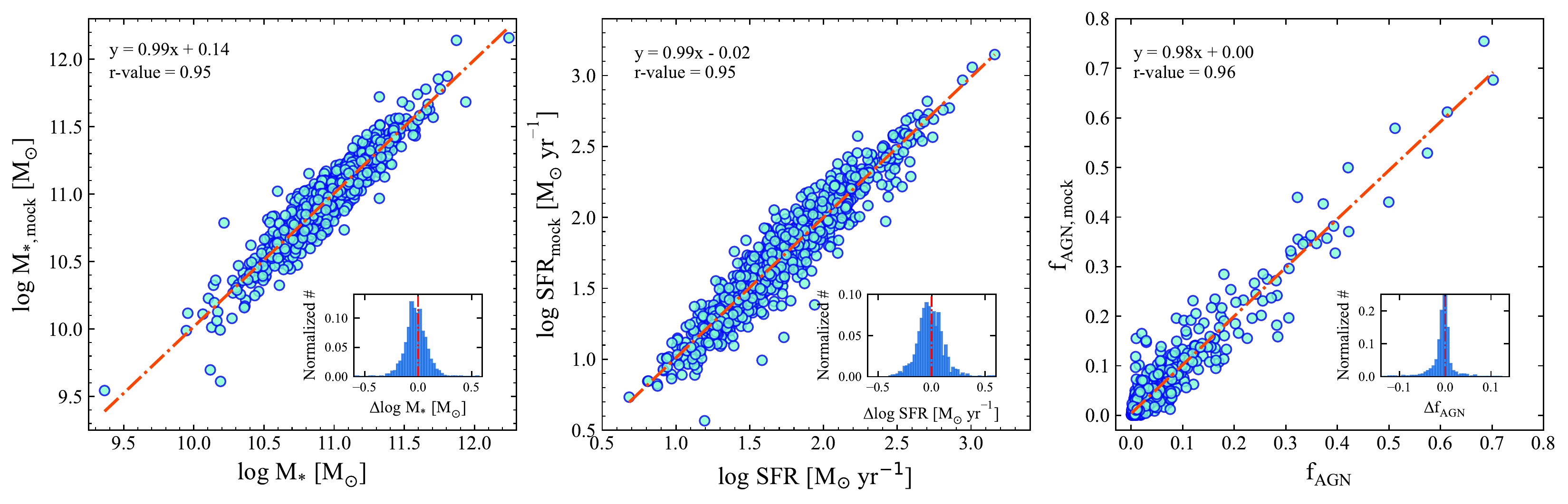}
\caption{The comparison between the true values (x-axis) and mock values (y-axis): stellar-mass ($left$), star formation rate ($middle$) and AGN fraction ($right$). The red line is the regression line with the equation and correlation coefficient given in the legend. The distribution of true minus estimated parameters has been shown in the insets.}
\label{fig:mock}
\end{figure}

\end{document}